\documentclass[conference,compsoc]{IEEEtran}

\ifCLASSOPTIONcompsoc
  \usepackage[nocompress]{cite}
\else
  \usepackage{cite}
\fi

\usepackage{amsmath}
\usepackage{amssymb}

\usepackage{graphicx}
\usepackage{tikz}

\usepackage{booktabs}
\usepackage{multirow}
\usepackage{array}
\usepackage{threeparttable}

\usepackage{xspace}
\usepackage{xcolor}
\usepackage{url}
\usepackage{hyperref}

\usepackage{enumitem}

\usepackage{titlesec}

\makeatletter
\renewcommand{\sectionautorefname}{\S\@gobble}
\renewcommand{\subsectionautorefname}{\S\@gobble}
\renewcommand{\subsubsectionautorefname}{\S\@gobble}
\makeatother

\graphicspath{{figures/}}

\newif\ifanonymous

\anonymousfalse

\newcommand{\sys}{{NRS\footnotesize{NOOPER}}\xspace}

\newcommand{\eg}{\textit{e.}\textit{g.}\xspace}

\newcommand{\nibf}[1]{\par\noindent\textbf{#1}\ }
\newcommand{\appref}[1]{\hyperref[#1]{Appendix~\ref*{#1}}}

\setlist[itemize]{noitemsep, topsep=0pt, leftmargin=*}

\title{\Large \sys: Enabling Real-Time Bidirectional Traffic Decoding in Commercial 5G NR Networks}

\ifanonymous

\author{
\IEEEauthorblockN{Anonymous Authors}
\IEEEauthorblockA{Anonymous Institution}
}

\else

\author{
\IEEEauthorblockN{Tuan Dinh Hoang\IEEEauthorrefmark{1}, Quang Huy Pham\IEEEauthorrefmark{1}, Taekkyung Oh\IEEEauthorrefmark{1}, Hazem Issa\IEEEauthorrefmark{1}, Marco Mezzavilla\IEEEauthorrefmark{2}, Yongdae Kim\IEEEauthorrefmark{1}}
\IEEEauthorblockA{\IEEEauthorrefmark{1}KAIST, Daejeon, Republic of Korea}
\IEEEauthorblockA{\IEEEauthorrefmark{2}Politecnico di Milano, Milan, Italy}
}

\fi

\begin{document}

\maketitle


\IEEEpeerreviewmaketitle


\begin{abstract}
\label{sec:abstract}
Decoding both uplink and downlink cellular traffic reveals lower-layer protocol behavior, unprotected signaling, and encrypted higher-layer packets beyond what scheduling metadata exposes. 
In commercial 5G NR networks, however, passive decoding must overcome UE-dedicated configurations hidden in encrypted signaling, unknown identifiers of ongoing connections, and the computational demands of wideband traffic.

We present \sys, a passive sniffer that decodes downlink control and bidirectional data traffic in real time, including traffic from already-connected UEs and traffic after encrypted configuration updates. 
\sys learns reusable physical-layer configurations through an instrumented open-source UE that reproduces commercial UE capabilities. 
It recovers connection-specific identifiers from ongoing transmissions and selects matching configurations through CRC-validated data decoding. 
A hybrid CPU--GPU pipeline sustains continuous wideband processing.

Evaluations on a controlled testbed and two commercial networks show that, compared with using only configurations from unencrypted signaling, \sys decodes 1.3$\times$ and 6.3$\times$ as many control messages and 9.0$\times$ and 17.7$\times$ as many downlink data blocks, respectively. 
\sys also decodes traffic from an already-connected UE during an ongoing phone call and achieves real-time bidirectional decoding on 100\,MHz commercial NR channels.

\end{abstract}

\section{Introduction}
\label{sec:introduction}

Although cellular networks encrypt higher-layer signaling and user data, radio transmissions still expose scheduling information and other lower-layer metadata to passive observers.
A software-defined-radio (SDR) sniffer can decode control and data channels to recover this metadata, the unprotected signaling, and encrypted higher-layer packets without breaking their encryption. 

Passive LTE sniffing has enabled identity mapping, user tracking,
and device fingerprinting~\cite{rupprecht2019breaking,hong2018guti,kotuliak2022ltrack,oh2023preventing},
as well as traffic fingerprinting and video identification despite
higher-layer encryption~\cite{rupprecht2019breaking,bae2022}.
Such observations support measurement
and security monitoring~\cite{tucker2025marlin,oh2026devilray}.
Many applications require more than decoded scheduling information:
for example, physical localization combines downlink timing information
with uplink signal measurements~\cite{kotuliak2022ltrack,oh2024enabling}.
They also require decoding in any connection state, since the target may already be connected when monitoring begins.
A practical sniffer must therefore meet three requirements:
(i) recover both downlink and uplink traffic,
(ii) decode throughout a connection, including after encrypted configuration updates and for already-connected user equipment (UE),
and (iii) operate in real time.

Extending these capabilities to 5G NR is challenging because decoding depends on substantially more UE-dedicated physical layer (PHY) configurations.
Unlike LTE, where a sniffer can test a few candidate parameters~\cite{hoang:ltesniffer}, NR makes even downlink control information (DCI) sizes and layouts depend on bandwidth parts (BWPs) and per-channel configurations delivered through Radio Resource Control (RRC) signaling.
Consequently, recovering DCI bits alone does not suffice: without the corresponding configuration, a sniffer cannot reliably interpret scheduling fields or decode the scheduled Physical Downlink Shared Channel (PDSCH) and Physical Uplink Shared Channel (PUSCH) transmissions.
NR also ties decoding to session-specific parameters, including the radio identifier (C-RNTI) and demodulation reference signal (DMRS) scrambling identities.

Obtaining these configurations raises three challenges.
First, the 5G base station (gNB) updates UE-dedicated PHY configurations through encrypted RRC Reconfiguration (C1).
Without these updates, a sniffer may use incorrect DCI
sizes or field layouts and cannot reliably decode 
scheduled data channels.
Second, even with the correct configuration, a sniffer that misses a UE's initial access lacks its C-RNTI and, in some networks, its DMRS scrambling identity (C2).
The unknown C-RNTI prevents DCI recovery and data descrambling, while an unknown DMRS identity prevents correct channel estimation.
Third, wideband, multi-UE decoding must keep pace with
traffic (C3).
A 100\,MHz NR carrier with 30\,kHz subcarrier spacing
requires processing 2,000 slots per second, including
candidate searches and bidirectional data decoding.

Existing 5G sniffers address only subsets of these challenges~(\autoref{tab:prior_work}).
5G-Sniffer~\cite{ludant20235g} and
5GDescrambler~\cite{windisch2026descrambler} recover
Physical Downlink Control Channel (PDCCH) scheduling information but do not provide general
PDSCH/PUSCH decoding.
NR-Scope~\cite{wan2024nrscope} and
SNI5GECT~\cite{luo2025sni5gect} decode data using
configurations obtained during connection establishment,
but do not recover configurations delivered through
encrypted RRC Reconfiguration or the missing session
context of already-connected UEs.
Consequently, existing research sniffers do not demonstrate
the combination of configuration recovery, already-connected
UE decoding, and real-time bidirectional processing of
ongoing commercial NR traffic.

We present \sys, a passive system that addresses these
three challenges in commercial 5G NR networks.
For C1, \sys exploits the observation that UEs with
similar radio capabilities often receive the same PHY
configurations.
It replays commercial UE capabilities and mutated variants
through an instrumented open-source UE with legitimate
SIM credentials, obtaining reusable configuration profiles
from its own RRC Reconfiguration messages.
During sniffing, \sys tests candidate profiles and identifies
matching ones through cyclic redundancy check (CRC)-validated transport-block decoding,
without knowing the target UE's exact capabilities.
For C2, \sys recovers the C-RNTIs of already-connected
UEs through syndrome-guided candidate ranking, control channel element (CCE)--RNTI
search-space pruning, and full polar decoding with CRC
validation.
It also recovers unknown PDSCH DMRS scrambling identities
through reference-signal correlation.
For C3, \sys integrates these techniques into a hybrid
CPU--GPU pipeline that parallelizes candidate searches
and processes bidirectional traffic across UEs
in real time.

We evaluated \sys on a controlled testbed and two commercial
5G standalone (SA) networks on different continents, Networks K and U.
In the testbed, \sys recovered 84--98\% of downlink and
uplink traffic under single-input single-output (SISO) and 75--98\% under $2\times2$ multiple-input multiple-output (MIMO), with receiver placements up to 18\,m.
By replaying 21 commercial UE capabilities and 83 variants,
\sys discovered 21 and 18 configuration profiles in the
two networks, increasing decoded downlink transport blocks
by 9.0$\times$ and 17.7$\times$, respectively.
Session-specific parameter recovery further enabled
bidirectional decoding of an already-connected UE during
an ongoing phone call.
The GPU-accelerated pipeline processed 100\,MHz commercial traffic in near real time
under moderate load, with about 120 active RNTIs per second.

Our contributions are as follows:
\begin{itemize}
    \item We develop capability-guided configuration profiling
    to recover reusable PHY configurations delivered through
    encrypted RRC signaling, without access to the target
    UE's capabilities or cryptographic keys.

    \item We enable decoding of already-connected UEs by
    combining syndrome-guided C-RNTI recovery, CCE-based
    candidate pruning, full polar decoding, and
    session-specific DMRS identity recovery.

    \item We implement \sys, an open-source, real-time
    bidirectional 5G NR sniffer with a hybrid CPU--GPU
    architecture for wideband, multi-UE decoding.

    \item We evaluate \sys on a controlled testbed and two
    commercial 5G SA networks, demonstrating decoding
    after encrypted configuration updates and during
    ongoing UE connections.
\end{itemize}

\section{Background}
\label{sec:preliminaries}

This section introduces the NR physical channels, their
decoding dependencies, and how UEs obtain the configurations
required for decoding. A UE communicates with its serving
gNB, which delivers radio configurations through broadcast
and RRC signaling.

\subsection{5G NR Physical Layer and Scheduling}
\label{subsec:5g_physical_layer}

\subsubsection{Time--frequency resources}
\begin{figure}[t]
\centering
\includegraphics[width=.8\columnwidth]{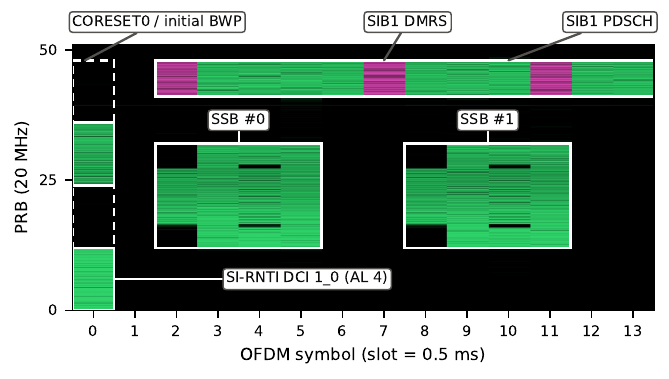}
\vspace{-3mm}
\caption{5G NR slot structure.}
\vspace{-3mm}
\label{fig:slot_arch}
\end{figure}

NR divides carrier bandwidth into physical resource blocks
(PRBs) of 12 subcarriers and time into slots of 14
orthogonal frequency-division multiplexing (OFDM) symbols.
At 30\,kHz subcarrier spacing (SCS), commonly used in
mid-band deployments, each slot lasts 0.5\,ms. 
\autoref{fig:slot_arch} illustrates an example NR
resource grid.
A BWP is a contiguous subset of carrier PRBs with its
own channel configurations. The initial BWP is cell-common
and supports random access, while the gNB configures
dedicated BWPs through RRC signaling. Multiple BWPs
may be configured, but only one per direction is active
at a time in each serving cell.

\subsubsection{Physical channels}
The PDCCH carries DCI, which schedules downlink and
uplink data transmissions on the PDSCH and PUSCH.
Unlike LTE's fixed control region, NR places the PDCCH
in configurable time--frequency regions called control
resource sets (CORESETs). CORESET0 is cell-common,
whereas additional CORESETs can be UE-dedicated.

PDCCH, PDSCH, and PUSCH use DMRS for channel
estimation. The Physical Uplink Control Channel (PUCCH)
carries uplink control information, while channel state
information (CSI) reference signals (CSI-RS) and sounding
reference signals (SRS) support downlink and uplink
channel measurements, respectively.

\subsubsection{Downlink Control Information}
DCI specifies resource allocation, modulation and coding
scheme (MCS), and retransmission information. Its CRC is scrambled with a 16-bit RNTI,
such as a UE's C-RNTI.

DCI formats 0\_0 and 0\_1 schedule PUSCH, while
1\_0 and 1\_1 schedule PDSCH. The fallback formats
0\_0 and 1\_0 have mostly fixed layouts, whereas the
non-fallback formats 0\_1 and 1\_1 depend on UE-dedicated
configurations. A DCI occupies one or more CCEs
according to its aggregation level (AL). Common search
spaces (CSSs) and UE-specific search spaces (USSs)
specify the PDCCH candidates and formats a UE monitors.

RRC configurations determine the sizes and interpretation
of many DCI fields. In DCI~1\_1, 18 of 26 fields have 
configuration-dependent widths~(\autoref{tab:dci11_fields_appendix} in~\appref{sec:app_dci_fields}).
For example, the BWP indicator depends on the number
of BWPs, the frequency-domain resource assignment (FDRA)
field on the BWP size and allocation type, and the
antenna-port field on the DMRS configuration. 
\autoref{tab:dci11_setup_vs_reconfig} shows the
three fields whose widths change in Network~K,
increasing the DCI size from 51 to 56 bits
(bit widths in parentheses).

\begin{table}[t]
\centering
\caption{DCI~1\_1 size changes in Network~K.}
\label{tab:dci11_setup_vs_reconfig}
\footnotesize
\setlength{\tabcolsep}{3pt}
\renewcommand{\arraystretch}{0.95}
\begin{tabular}{@{}lcc@{}}
\hline
Field & RRC Setup & RRC Reconf. \\
\hline
BWP indicator & 1 BWP (1) & 2 BWPs (2) \\
FDRA (273 PRBs) & Type 1 (16) & dynamicSwitch (19) \\
Antenna ports & DMRS type 1 (4) & DMRS type 2 (5) \\
\hline
\textbf{Total DCI bits} & \textbf{51} & \textbf{56} \\
\hline
\end{tabular}
\end{table}

\subsection{Standard DCI, PDSCH, and PUSCH Decoding}
\label{subsec:decoding_chains}

\begin{figure}[t]
\centering
\includegraphics[width=.8\columnwidth]{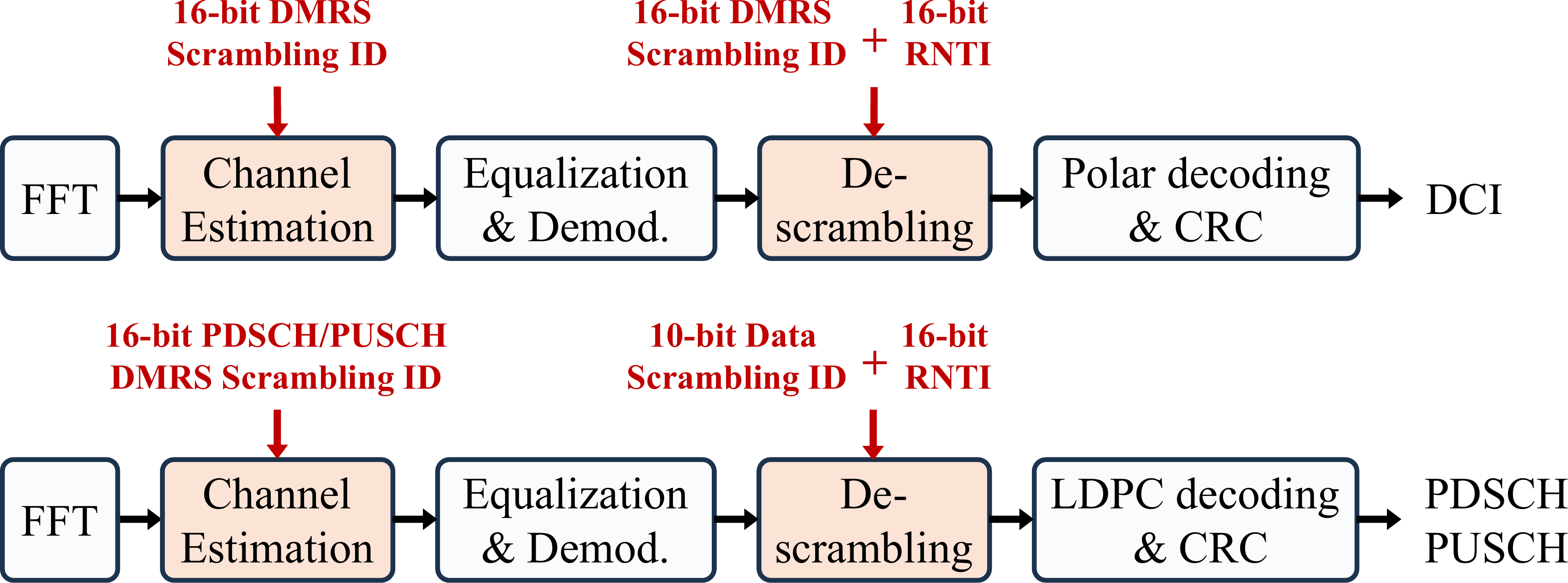}
\caption{Standard decoding procedures of DCI and PDSCH/PUSCH. FFT and LDPC denote fast Fourier transform and low-density parity-check.}
\label{fig:decoding_chains}
\end{figure}

\autoref{fig:decoding_chains} illustrates the parameters
required for DCI and PDSCH/PUSCH decoding.
Beyond cell-common configurations, DCI recovery requires the PDCCH DMRS scrambling identity from the UE-dedicated configuration and the session-specific C-RNTI.
Data decoding further depends on the corresponding DMRS and data-scrambling identities, some of which can also be session-specific.
Missing or incorrect parameters can prevent channel
estimation or descrambling, while a valid DCI CRC does
not guarantee correct field interpretation or successful
data decoding.

\subsection{UE Configuration-Oriented States}
\label{subsubsec:ue_states}

\begin{figure}[t]
\centering
\includegraphics[width=.8\columnwidth]{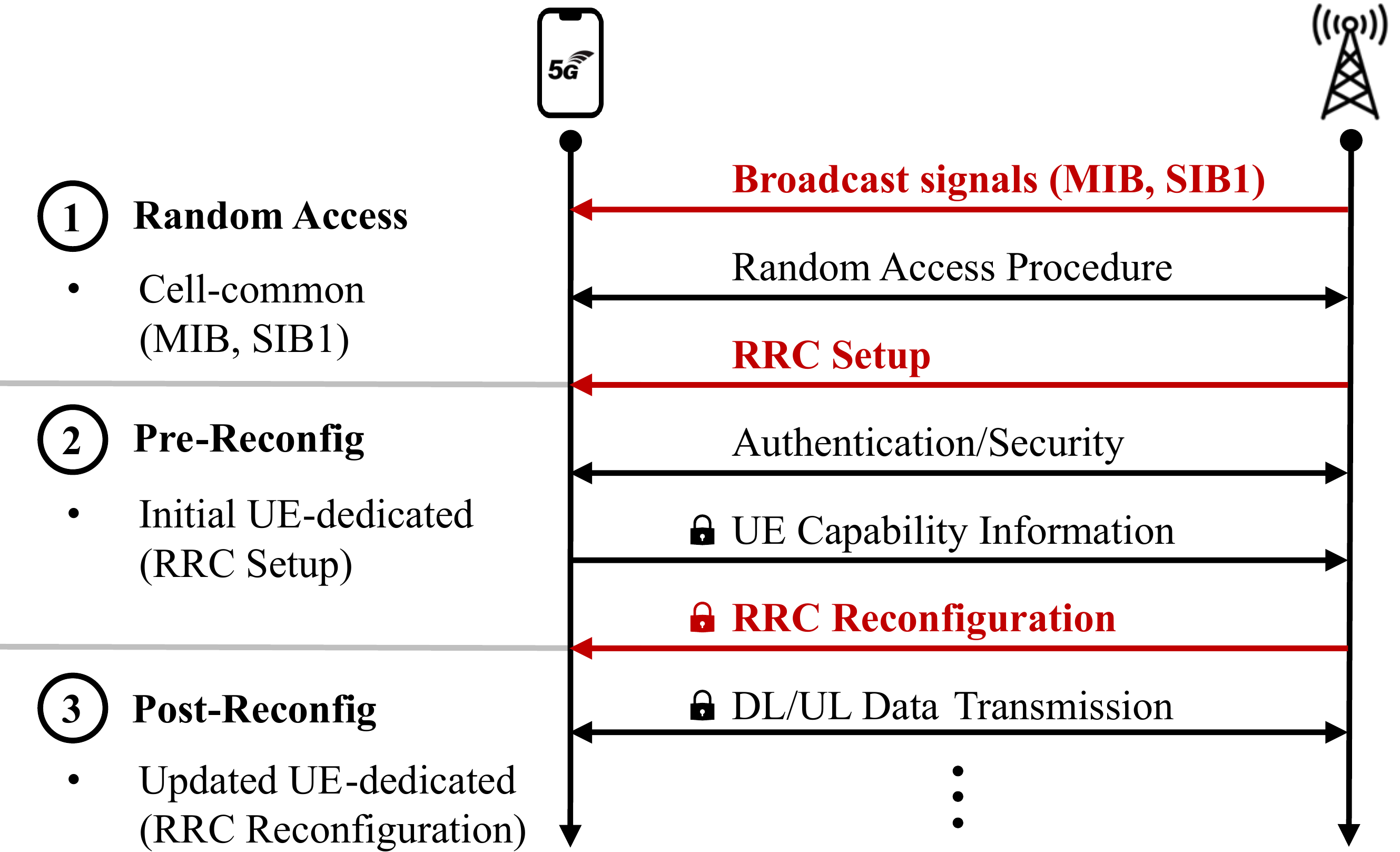}
\caption{UE configuration states and associated signaling.}
\label{fig:ue_states}
\end{figure}

We distinguish three configuration-oriented UE states
according to the signaling through which decoding
configurations become available~(\autoref{fig:ue_states}).

\nibf{RA state.}
During random access (RA), the UE obtains cell-common
configurations through the synchronization signal block
(SSB), master information block (MIB), and system
information block 1 (SIB1). These provide CORESET0
and the initial BWP. The random access response (RAR)
assigns a temporary C-RNTI, which becomes the UE's
C-RNTI after contention resolution.

\nibf{Pre-Reconf state.}
After random access, unencrypted RRC Setup provides
the initial UE-dedicated configurations, including BWPs,
CORESETs, search spaces, and per-channel parameters.
The UE subsequently activates security and reports
its radio capabilities.

\nibf{Post-Reconf state.}
After security activation, the gNB delivers encrypted
RRC Reconfiguration messages that update UE-dedicated
configurations based on the reported capabilities.
These updates may change BWPs, resource-allocation types,
and MCS tables. Data radio bearers are also established,
enabling user-plane traffic.

\subsection{Configuration Classification}
\label{subsubsec:config_classification}
Decoding parameters differ in their scope and reusability.
We classify them into three groups.

\nibf{Cell-common configuration.}
Broadcast through the MIB and SIB1, it includes CORESET0
and the initial BWP shared by UEs in the same cell.

\nibf{UE-dedicated configuration.}
Delivered through RRC Setup and RRC Reconfiguration,
it includes dedicated BWPs and per-channel configurations.
A gNB may assign identical PHY configurations to different
UEs based on their radio capabilities. We call a reusable
set of such configurations a configuration profile.

\nibf{Session-specific parameters.}
These parameters vary between connections and cannot
be directly reused. Examples include the C-RNTI assigned
during random access and the PDSCH DMRS scrambling
identity assigned per connection, as observed in Network~U~(\autoref{subsubsec:dmrs_id_recovery}).
Which parameters vary by session depends on the gNB.

\subsection{Passive Sniffing Procedure}
\label{subsec:passive_sniffer_procedure}

Passive decoding follows four dependent steps.
In Step (1), the sniffer blind-decodes PDCCH candidates
using the configured BWP, CORESET, search spaces,
and DCI sizes.
In Step (2), it unpacks the recovered DCI bits according
to configuration-dependent field layouts.
In Step (3), it interprets the fields using resource-allocation
rules, MCS tables, and time-domain allocation tables.
In Step (4), it decodes the scheduled PDSCH or PUSCH
using the corresponding channel, DMRS, and scrambling
parameters.

\section{Motivation and Challenges}
\label{sec:motivation}
We consider an adversary who uses a passive sniffer to decode the uplink and downlink traffic of multiple UEs in a commercial 5G NR cell, including UEs that are already connected. 
Such a sniffer faces three challenges.

An adversary may use NR-Scope~\cite{wan2024nrscope} or
SNI5GECT~\cite{luo2025sni5gect} to decode traffic during
connection establishment. These sniffers obtain cell-common
configurations from MIB/SIB1 and initial UE-dedicated
configurations from unencrypted RRC Setup, but cannot
observe subsequent updates in encrypted RRC
Reconfiguration. In Network~K, such an update changes
the DCI~1\_1 size from 51 to 56 bits
(\autoref{tab:dci11_setup_vs_reconfig}).
A sniffer retaining the old configuration therefore searches
for the wrong DCI size. Trying additional sizes alone is
insufficient: even with a valid DCI CRC, an incorrect
configuration can misinterpret scheduling fields, while
PDSCH/PUSCH decoding additionally requires the correct
MCS tables and DMRS parameters.
Thus, \textbf{C1 is to infer UE-dedicated PHY configurations
delivered through encrypted RRC Reconfiguration without
decrypting it.}

Even if the adversary obtains the updated configuration
(C1), monitoring may begin after the UE has connected.
The sniffer then misses session-specific parameters
conveyed during connection establishment, notably the
C-RNTI and, in some networks, the PDSCH DMRS scrambling
identity. Without the C-RNTI, DCI CRC validation and
data descrambling fail; an unknown DMRS identity also
prevents correct PDSCH channel estimation.
Although 5G-Sniffer~\cite{ludant20235g} and
5GDescrambler~\cite{windisch2026descrambler} recover
RNTIs from ongoing PDCCH transmissions, they do not
decode the corresponding PDSCH and PUSCH traffic.
Thus, \textbf{C2 is to recover the session-specific parameters
of already-connected UEs from ongoing transmissions.}

Even with the required configurations and session-specific
parameters, real-time bidirectional decoding faces five
additional challenges:

\begin{itemize}
    \item \textbf{Wideband processing.}
    A 100\,MHz NR carrier with 30\,kHz SCS generates
    2,000 slots per second, requiring continuous
    signal processing and channel decoding.

    \item \textbf{Multi-UE decoding.}
    Each slot may contain transmissions from multiple
    UEs, requiring parallel decoding and repeated
    searches over DCI and RNTI candidates.

    \item \textbf{Cross-slot scheduling.}
    Unlike LTE FDD's fixed 4\,ms delay, an NR DCI may schedule PUSCH within 1\,ms, leaving little time to prepare for PUSCH decoding.

    \item \textbf{Uplink synchronization.}
    UEs have different propagation delays and uplink
    timing offsets, requiring UE-specific synchronization
    for reliable PUSCH decoding.

    \item \textbf{Dynamic UE management.}
    UEs connect and disconnect continuously, while
    C-RNTIs may be reassigned. The sniffer must track each UE's decoding context.
\end{itemize}

Thus, \textbf{C3 is to sustain real-time bidirectional
decoding across multiple UEs while handling these
timing, synchronization, and computational demands.}

Table~\ref{tab:prior_work} summarizes these gaps.
We next present \sys, which addresses C1--C3.

\begin{table}[t]
\centering
\caption{Comparison of 5G passive sniffers.}
\label{tab:prior_work}
\footnotesize
\setlength{\tabcolsep}{4pt}
\begin{tabular}{@{}lcccccccc@{}}
\hline
Sniffer & Ctrl & DL & UL & Pre & Post & Conn & RT & BW \\
\hline
5G-Sniffer~\cite{ludant20235g}
  & \checkmark & $\times$ & $\times$
  & R & R & R & DCI & 20 \\
5GDescrambler~\cite{windisch2026descrambler}
  & \checkmark & $\times$ & $\times$
  & R & R & R & DCI & 20 \\
NR-Scope~\cite{wan2024nrscope}
  & \checkmark & \checkmark & $\times$
  & \checkmark & $\times$ & $\times$ & Pre & 100 \\
SNI5GECT~\cite{luo2025sni5gect}
  & \checkmark & \checkmark & \checkmark
  & \checkmark & $\times$ & $\times$ & Pre & 50 \\
Golden Sniffer~\cite{palama2026golden}
  & \checkmark & \checkmark & $\times$
  & \checkmark & $\times$ & $\times^{*}$ & $\times$ & $\times$ \\
\hline
\textbf{NRSNOOPER}
  & \checkmark & \checkmark & \checkmark
  & \checkmark & \checkmark & \checkmark & All & 100 \\
\hline
\end{tabular}
\vspace{2pt}

\parbox{\columnwidth}{\scriptsize
Ctrl: PDCCH; DL/UL: PDSCH/PUSCH;
Pre/Post: configuration state;
Conn: already-connected UEs;
R: raw DCI bits without interpretation;
RT: real-time decoded traffic
(DCI, Pre-Reconf, or All; $\times$: offline only);
BW: bandwidth in MHz;
$^{*}$: only while the gNB uses fallback DCIs 0\_0/1\_0, mostly in open-source gNBs.
}
\end{table}

\section{Overview}
\label{sec:overview}

\sys targets real-time PDCCH, PDSCH, and PUSCH decoding
for multiple UEs in both open-source and commercial 5G NR
networks, including Post-Reconf traffic (C1) and
already-connected UEs (C2).
Its coverage depends on received signal quality,
particularly for uplink transmissions from distant UEs.

\subsection{Threat Model}
\label{subsec:threat_model}

We consider an adversary who aims to decode the
bidirectional traffic of UEs in a commercial 5G NR cell.
The adversary operates in two phases.

\nibf{Phase 1: Configuration profiling (active).}
Before sniffing, the adversary connects its own open-source
UE with a valid commercial SIM to the target network,
replays commercial UE capabilities and their variants,
and observes the RRC configurations assigned to its
own UE~(\autoref{sec:design}).

\nibf{Phase 2: Traffic sniffing (passive).}
The adversary captures and decodes over-the-air traffic
using SDRs and a commodity PC, without transmitting
signals or interacting with the observed UEs or gNB.

The adversary has no access to target UEs, network
infrastructure, privileged interfaces, or cryptographic
keys, and cannot decrypt protected signaling.

\subsection{Approaches}
\label{subsec:challenges_approaches}

\sys addresses C1--C3 through capability-guided
configuration profiling (A1), session-specific parameter
recovery (A2), and real-time multi-UE decoding (A3).

\nibf{A1.1: COTS UE Capability Collection and Mutation.}
Because the gNB assigns PHY configurations based on
reported UE capabilities, a configuration obtained from
one UE may not apply to others. \sys collects capabilities
from commercial smartphones and mutates PHY-relevant
fields to cover a broader range of configurations~(\autoref{subsubsec:capability_mutation}).

\nibf{A1.2: Open-Source UE-based Configuration Discovery.}
\sys replays these capabilities through an instrumented
OpenAirInterface (OAI) UE with a valid commercial SIM.
By observing its own decrypted RRC Reconfiguration
messages, \sys identifies reusable configuration profiles
and parameters that vary across connections~(\autoref{subsubsec:ue_config_discovery}).

To address C2, \sys recovers C-RNTIs from ongoing PDCCH
transmissions (A2.1--A2.2) and, when necessary,
session-specific PDSCH DMRS scrambling identities
through reference-signal correlation~(\autoref{subsubsec:dmrs_id_recovery}).

\nibf{A2.1: Syndrome-Guided C-RNTI Recovery.}
Building on prior work~\cite{gardner2020efficient,richards2024rnti},
\sys ranks RNTI hypotheses using polar frozen-bit
syndromes and performs full soft-decision polar decoding
with CRC validation on the highest-ranked candidates.

\nibf{A2.2: CCE-based RNTI Candidate Pruning.}
\sys further prunes RNTI hypotheses using the specification-defined relationship between a C-RNTI and its permitted PDCCH CCE locations, excluding those inconsistent with the observed location before syndrome matching~(\autoref{subsec:recovering_session_parameters}).

\nibf{A3: Real-Time Multi-UE Decoding.}
\sys uses a hybrid CPU--GPU pipeline to process wideband
traffic and decode multiple UEs in parallel. The GPU
accelerates PDCCH candidate searches, C-RNTI recovery,
and DMRS-identity correlation, while the CPU handles
cross-slot scheduling, adaptive uplink synchronization,
and active-UE management. Together, these mechanisms
address the timing and computational demands of C3~(\autoref{subsec:design_realtime}
and~\autoref{subsec:eval_realtime}).

\subsection{Overall Architecture}
\label{subsec:overview_architecture}

\begin{figure}[t]
\centering
\includegraphics[width=\columnwidth]{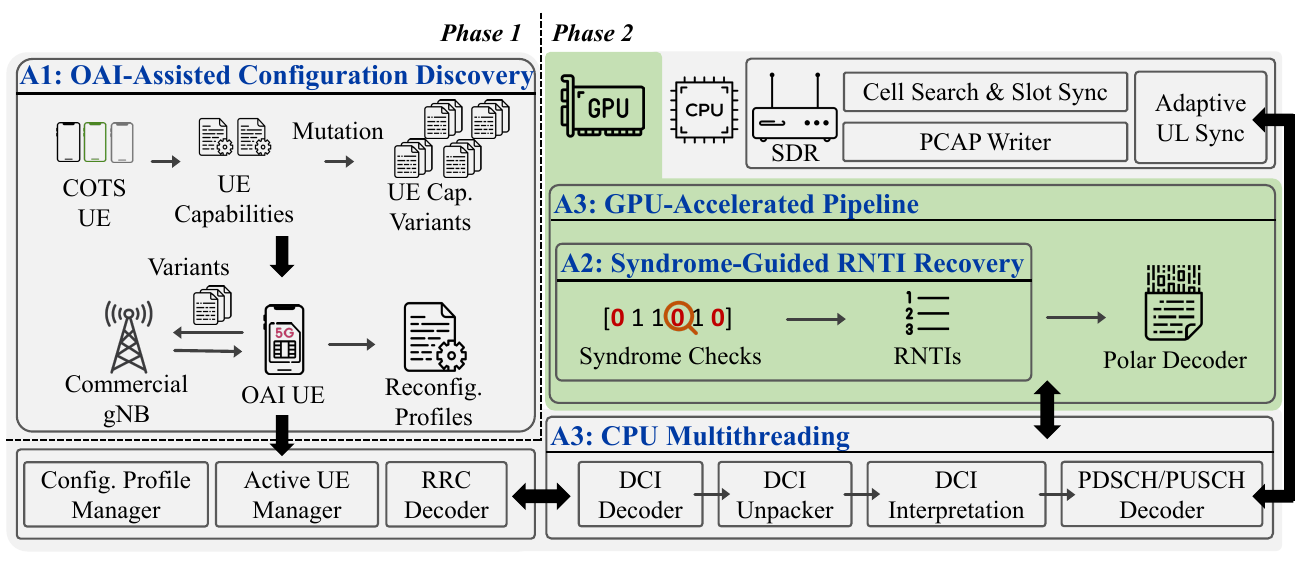}
\caption{Design overview of \sys.}
\label{fig:design_overview}
\end{figure}

\autoref{fig:design_overview} summarizes the two phases
of \sys. Phase~1 discovers reusable configuration profiles
and identifies session-specific parameters through the
adversary's own UE (A1). Phase~2 passively decodes
traffic using these profiles and session parameters
obtained from observed signaling or recovered from
ongoing transmissions (A2). During sniffing, \sys
obtains cell-common configurations from MIB and SIB1.

\nibf{Newly connecting UEs.}
\sys obtains the C-RNTI from the RAR and the initial
UE-dedicated configuration and session-specific parameters
from RRC Setup. For Post-Reconf traffic, it tests candidate
profiles and identifies the matching one through
PDSCH/PUSCH CRC validation.

\nibf{Already-connected UEs.}
Having missed initial access, \sys derives candidate DCI
sizes and search spaces from the profiles and recovers
the C-RNTI from ongoing PDCCH transmissions (A2).
It then interprets DCI using candidate profiles and,
when necessary, recovers the session-specific PDSCH
DMRS scrambling identity through reference-signal
correlation. A CRC-valid transport block confirms the
matching profile and scrambling identity.

A1 and A2 are complementary: profiles enable
configuration-aware DCI interpretation, while recovered
session parameters enable DCI and data decoding.
A3 integrates these operations into a real-time
CPU--GPU pipeline~(\autoref{subsec:design_realtime}).
\section{Design}
\label{sec:design}

We organize the design around the architecture in~\autoref{subsec:overview_architecture}.

\subsection{OAI-based RRC Reconfiguration Discovery}
\label{subsec:rrc_reconfig_discovery}
\sys discovers the configurations hidden in encrypted RRC Reconfiguration (C1) by connecting an OAI UE~\cite{openairinterface5g} that reports diverse UE capabilities to the network.
It groups these configurations into profiles and reuses them to decode other UEs with similar capabilities (\autoref{subsubsec:config_classification}).

\subsubsection{Collecting and Mutating COTS UE Capabilities}
\label{subsubsec:capability_mutation}

\nibf{Extracting COTS UE Radio Capability Information.}
We collect UE Capability Information from 21 COTS UEs
spanning five phone vendors and five baseband vendors.
\appref{sec:app_ue_caps} lists the devices~(\autoref{tab:app_cots}).
We obtain each capability either by connecting the UE
to an open-source gNB in the target band and
capturing its capability in a PCAP file, or by connecting
it to the target network with a commercial SIM and
extracting the message from an XCAL trace~\cite{xcal}.
We decode each message into XML, which serves as a
mutation baseline loaded by the OAI UE
(\autoref{subsubsec:ue_config_discovery}).

\nibf{Mutating UE Radio Capabilities.}
The 21 COTS UEs may not cover all capabilities reported
by UEs in the target network. We therefore generate
variants by mutating fields that affect PHY configurations
or DCI layouts, based on their semantics in TS~38.306~\cite{3gpp38306}
and TS~38.331~\cite{3gpp-5g-rrc}.
For example, \texttt{pdsch-256QAM-FR1} and
\texttt{pusch-256QAM} determine the MCS table;
\texttt{maxNumberMIMO-LayersPDSCH} and
\texttt{maxNumberMIMO-LayersCB-PUSCH} constrain the
number of MIMO layers, with the latter also changing the width
of the precoding field in DCI~0\_1; and
\texttt{dynamicSwitchRA-Type0-1-PDSCH} determines
whether the gNB dynamically switches resource allocation
types 0 and 1, changing the width of the frequency-domain
resource assignment field in DCI~1\_1.
We use an LLM to extract field semantics and suggest candidate fields that may affect
the PHY configurations in RRC Reconfiguration, then
verify each suggestion against the specifications.
Each variant deletes a field, adds one copied from another
phone, or changes its value, either individually or in
small groups that must remain consistent (\eg, fields
describing a second Tx chain).
We also add band combinations, including intra-band
non-contiguous carrier aggregation (CA), to trigger CA.
In total, we mutate 57 fields and several band combinations,
generating 83 variants covering BWP, cross-slot scheduling,
resource allocation, MIMO, modulation, CSI, DMRS, and
PUCCH capabilities.
\appref{sec:app_ue_caps} lists the complete set of mutated fields~(\autoref{tab:app_cap_fields}).

\subsubsection{Connecting the OAI UE to Commercial Networks}
\label{subsubsec:ue_config_discovery}

\nibf{Enabling OAI UE to Connect to Commercial Networks.}
The default OAI UE connects to open-source gNBs but not
commercial networks. We extend it to access a commercial
SIM through the PC/SC interface and complete the
Authentication and Key Agreement (AKA) procedure with
its credentials.
Because some networks reject the default international mobile equipment identity (IMEI) or limited
radio capabilities, we make the IMEI configurable and
allow the OAI UE to load and report capabilities from XML
files, even when they exceed its supported PHY features.
We also make the access point name (APN) and IP address type (IPv4 or IPv6)
configurable and disable unsupported CSI reports.
Finally, the OAI UE captures the established RRC security
context and logs its decrypted RRC messages, including
RRC Reconfiguration.

\nibf{Replaying Capabilities.}
For each connection, the OAI UE follows the standard
registration procedure: after RRC Setup, it sends a
Registration Request, completes AKA and security
activation, and responds to the UE Capability Enquiry
with the loaded capability.
The gNB then configures the OAI UE through RRC
Reconfiguration reflecting that capability.
We replay all 104 capabilities (21 COTS capabilities
and 83 variants) in separate connections.
To limit network load, we attempt at most three
connections per capability and mark it as failed if none
receives an RRC Reconfiguration.
For each successful connection, we save the reported
capability, RRC Setup, and decrypted RRC Reconfiguration
in a PCAP file.

\subsubsection{Extracting PHY Configuration Changes}
\label{subsubsec:phy_config_extraction}
\sys then compares each connection's RRC Reconfiguration with its RRC Setup and extracts the changed per-channel configurations (\eg, PDCCH, PDSCH, and PUSCH).
For each capability, we compare the resulting configurations across repeated connections. We treat parameters that remain stable across connections as reusable UE-dedicated configurations, and those that vary between connections as session-specific parameters. 
We use an LLM to help identify candidate differences between the messages and verify each against them.
Finally, \sys groups capabilities receiving the same
UE-dedicated PHY configuration into configuration profiles.
\appref{sec:app_profiles} lists the resulting profiles in Network~K and Network~U.

\subsubsection{Profiles and Session-Specific Parameters}
\label{subsubsec:session_param_discovery}
Except for the PDSCH DMRS scrambling identity in Network~U, both networks assign the same initial UE-dedicated PHY configuration in RRC Setup across all tested capabilities and connections.
\sys can therefore reuse this configuration
for Pre-Reconf UEs whose RRC Setup it did not observe.
Because the gNB considers only a subset of capability fields, Network~K and Network~U
assign 21 and 18 distinct profiles, respectively
(\autoref{subsec:eval_profiling}).

Comparing connections also reveals the session-specific
parameters that \sys must recover during sniffing.
Both networks assign a new temporary C-RNTI (TC-RNTI) in each RAR, which
becomes the UE's C-RNTI after successful contention
resolution. Network~U also assigns a different PDSCH
DMRS scrambling identity to each connection in RRC
Setup and retains it after RRC Reconfiguration. 
\sys therefore recovers this identity during sniffing
(\autoref{subsubsec:dmrs_id_recovery}) rather than storing
it in a profile. 
Although the specifications allow per-UE PDCCH DMRS, PDSCH/PUSCH data, and PUSCH DMRS scrambling identities~\cite{3gpp-5g-rrc}, both networks keep them identical across connections, so \sys stores them in the profiles.

\subsection{Recovering Session-Specific Parameters}
\label{subsec:recovering_session_parameters}
While UE-dedicated configurations are shared by UEs with similar capabilities and thus need to be discovered only once (\autoref{subsec:rrc_reconfig_discovery}), session-specific parameters vary across connections and must be recovered during sniffing (\autoref{subsubsec:session_param_discovery}).
We identify two such parameters: the C-RNTI in both networks
and the PDSCH DMRS scrambling identity in Network~U.

\subsubsection{Frozen-Bit Syndrome-Guided C-RNTI Recovery}
\label{subsubsec:syndrome_rnti_recovery}

\sys recovers the C-RNTI of an already-connected UE using
the frozen-bit syndrome technique of prior work
\cite{gardner2020efficient,richards2024rnti}.
We formalize the key relations underlying this technique.

In PDCCH encoding, a DCI payload of $A$ bits is appended
with a 24-bit CRC, yielding $K=A+24$ information bits.
After input interleaving, these bits occupy the information
positions of a length-$N$ polar encoder input $\mathbf{u}$;
the remaining $N-K$ positions are frozen to zero.
Polar encoding produces $\mathbf{x}=\mathbf{u}G_N$,
where $G_N$ is the polar generator matrix over
$\mathrm{GF}(2)$.
Let $\mathcal{F}$ denote the frozen positions.
We define the frozen-bit syndrome function as
\begin{equation}
    f(\mathbf{v})=\bigl(\mathbf{v}G_N^{-1}\bigr)_{\mathcal{F}},
    \label{eq:frozen_syndrome}
\end{equation}
where all operations are over $\mathrm{GF}(2)$.
Since frozen positions are zero, every valid polar codeword
satisfies $f(\mathbf{x})=\mathbf{0}$.

For UE-specific PDCCHs, coded-bit scrambling is initialized as
\begin{equation}
    c_{\mathrm{init}}=(r\cdot 2^{16}+n_{\mathrm{ID}})\bmod 2^{31},
    \label{eq:pdcch_scrambling_init}
\end{equation}
where $r$ is the unknown C-RNTI and $n_{\mathrm{ID}}$ is the
PDCCH scrambling identity.
Our OAI-UE-based discovery shows that $n_{\mathrm{ID}}$
is shared by UEs within each CORESET and known from its
configuration profile. Thus, the scrambling sequence
$\mathbf{c}(r)$ depends only on the RNTI hypothesis.

For clarity, we first derive the syndrome relation for
a PDCCH codeword before rate matching. Let
$\hat{\mathbf{b}}$ denote hard decisions on its received
scrambled bits:
\begin{equation}
    \hat{\mathbf{b}}=\mathbf{x}\oplus\mathbf{c}(r^\star)\oplus\mathbf{e},
    \label{eq:received_scrambled_pdcch}
\end{equation}
where $r^\star$ is the transmitted C-RNTI and $\mathbf{e}$
represents hard-decision errors.
Applying $f$ and using $f(\mathbf{x})=\mathbf{0}$ gives
\begin{equation}
    f(\hat{\mathbf{b}})=f\bigl(\mathbf{c}(r^\star)\bigr)\oplus f(\mathbf{e}).
    \label{eq:syndrome_separation}
\end{equation}
Define $H=f(\hat{\mathbf{b}})$ as the received syndrome and
$S(r)=f(\mathbf{c}(r))$ as the signature of an RNTI hypothesis.
Then $H=S(r^\star)\oplus f(\mathbf{e})$.
For any hypothesis $r$, linearity further gives
\begin{equation}
    f\bigl(\hat{\mathbf{b}}\oplus\mathbf{c}(r)\bigr)=H\oplus S(r).
    \label{eq:syndrome_lookup}
\end{equation}

The above derivation assumes that the received scrambled
bits directly correspond to the length-$N$ polar codeword.
In the actual implementation, \sys first converts the
received PDCCH log-likelihood ratios (LLRs) to hard decisions and applies inverse
rate matching to map the bits back to their polar codeword
positions before computing the syndrome.
For punctured codewords, it uses only frozen-bit constraints
that do not involve punctured positions.
For repeated codewords, it uses only the first received
copy.
For shortened codewords, it zero-fills the shortened positions, as the encoder does, so all frozen-bit constraints remain usable.
The resulting syndrome is then compared against
the precomputed RNTI signatures.
\appref{sec:app_syndrome} provides
the complete derivation and details of these transformations.

Eq.~\eqref{eq:syndrome_lookup} separates computations
dependent on the received PDCCH candidate from those
dependent on the RNTI hypothesis.
\sys precomputes $S(r)$ for each PDCCH configuration
and computes $H$ once per received candidate.
Moreover, the modulo-$2^{31}$ operation in
Eq.~\eqref{eq:pdcch_scrambling_init} makes the most
significant bit of the 16-bit C-RNTI irrelevant:
$\mathbf{c}(r)=\mathbf{c}(r\oplus 2^{15})$.
Thus, \sys evaluates only $2^{15}$ distinct signatures.
\sys ranks the hypotheses by their Hamming distance
$d(r)=\operatorname{wt}(H\oplus S(r))$, replacing
a soft-decision polar decoding per hypothesis with
an XOR and a population count.
Because hard-decision errors perturb $H$, the correct
C-RNTI may not have the smallest distance.
\sys therefore retains the 16 closest hypotheses and runs soft-decision polar decoding
with CRC validation on each, exploiting bit reliability
to correct errors.

\subsubsection{Reducing the Number of RNTI Candidates}
\label{subsubsec:rnti_candidate_reduction}

Although syndrome matching is cheaper than full polar
decoding, evaluating all $2^{15}$ signatures for every
PDCCH candidate remains expensive.
\sys further prunes RNTI hypotheses using the
C-RNTI-dependent mapping of PDCCH candidates to CCE locations in a UE-specific search
space.
Without cross-carrier scheduling, the starting CCE index
$j$ of candidate $m$ at aggregation level $L$ is
\cite{3gpp_ts_38213}
\begin{equation}
\label{eq:rnti_cce_mapping}
 j = L\!\left[\left(Y_{p,n}(r) +
 \left\lfloor\frac{mN_{\mathrm{CCE},p}}{L M^{(L)}}\right\rfloor\right)
 \bmod\left\lfloor\frac{N_{\mathrm{CCE},p}}{L}\right\rfloor\right],
\end{equation}
where
\begin{equation}
\label{eq:rnti_search_space_hash}
 Y_{p,n}(r) = A_p^{n+1}r\bmod 65537.
\end{equation}
Here, $N_{\mathrm{CCE},p}$ is the number of CCEs in
CORESET $p$; $M^{(L)}$ is the number of candidates at
aggregation level $L$; $m$ is the candidate index;
$n$ is the slot index within the radio frame; and
$A_p\in\{39827,39829,39839\}$ depends on $p\bmod 3$.
The profile provides the CORESET and search-space
parameters, while the observed candidate provides its
starting CCE, slot, and aggregation level.
\sys therefore retains only RNTIs satisfying
Eq.~\eqref{eq:rnti_cce_mapping} for at least one
configured candidate index $m$.

The mapping distributes the $M^{(L)}$ candidates of
each RNTI among distinct blocks of $L$ CCEs, out of
$B=\lfloor N_{\mathrm{CCE},p}/L\rfloor$ blocks.
Because 65537 is prime and $A_p$ is nonzero modulo
65537, $Y_{p,n}(\cdot)$ permutes the residues modulo
65537. 
Consequently, RNTI hypotheses spread nearly uniformly across the $B$ blocks.
Thus, about a fraction $s=M^{(L)}/B$ of RNTIs 
can use a given CCE location, and A2.2 prunes the remaining $1-s$.
This fraction depends only on $N_{\mathrm{CCE},p}$,
$L$, and $M^{(L)}$, so \sys precomputes the surviving
RNTIs for each slot and block.

For a 45-CCE CORESET, pruning ranges from 97.8\%
for a single candidate at $L{=}1$ to zero when
$M^{(L)}\geq B$.
Both commercial networks use a 45-CCE CORESET.
Network~K uses $L\in\{4,8\}$ with
$M^{(L)}=\{6,2\}$, pruning 45\% and 60\% of RNTIs.
Network~U mostly uses $L\in\{1,2,4,8\}$ with
$M^{(L)}=\{4,2,2,1\}$, pruning 80--91\%.
Since each signature is shared by two RNTIs
(\autoref{subsubsec:syndrome_rnti_recovery}),
\sys retains a signature if either RNTI survives.
Weighted by the aggregation levels observed on air,
A2.2 removes 24\% of signatures in Network~K and
79\% in Network~U, consistent with the measured
22\% and 81\% reductions in tested hypotheses.

\subsubsection{Recovering PDSCH DMRS Scrambling Identity}
\label{subsubsec:dmrs_id_recovery}
Network~U assigns each connection a different PDSCH DMRS scrambling identity in RRC Setup (\autoref{subsubsec:session_param_discovery}).
For an already-connected UE whose RRC Setup predates
sniffing, \sys recovers this identity through DMRS
correlation.
For each grant of a UE with an unknown identity,
\sys extracts the received DMRS symbols at the resource
elements specified by the decoded DCI and correlates
them with pre-generated sequences for candidate identities.
It validates the best-matching identity through PDSCH decoding and, if the transport-block CRC passes, stores it under the UE's C-RNTI in the active-UE table for subsequent PDSCH transmissions.

\subsection{Configuration-aware End-to-End Decoding}
\label{subsec:configuration_aware_end_to_end_decoding}

\sys decodes each UE's traffic using the discovered
configuration profiles (\autoref{subsec:rrc_reconfig_discovery})
and recovered session-specific parameters
(\autoref{subsec:recovering_session_parameters}).

\subsubsection{Cell Search and Initial Configurations}
\label{subsubsec:cell_search_sync}

Unlike LTE, NR allows the SSB to occupy different
synchronization raster positions within a carrier.
\sys therefore searches for the SSB in time and
frequency, decodes MIB and SIB1, and derives the
downlink slot boundary from the SSB.
Using the cell-common configuration from SIB1,
\sys decodes the RA procedure, including
the RAR and RRC Setup. 
The RAR provides the C-RNTI, and RRC Setup provides the initial UE-dedicated configuration and, in Network~U, the PDSCH DMRS scrambling identity.
For Post-Reconf traffic, \sys uses the discovered
configuration profiles
(\autoref{subsec:rrc_reconfig_discovery}).

\subsubsection{PDCCH and DCI Decoding}

\sys derives PDCCH candidates from the CORESET and
search-space configurations obtained through broadcast
signaling, RRC Setup, or the discovered profiles.
It handles three cases: CSSs, USSs without a configured
PDCCH DMRS scrambling identity, and USSs with one.
CSSs use cell-common configurations for system
information, paging, and random access, whereas USSs
use UE-dedicated configurations. For USSs without a
configured PDCCH DMRS scrambling identity, the physical
cell identity (PCI) is used~\cite{3gpp38211}.
Otherwise, as in Networks~K and~U, PDCCH decoding
depends on the configured scrambling identity and
C-RNTI (Eq.~\eqref{eq:pdcch_scrambling_init}).
\sys recovers candidate C-RNTIs
(\autoref{subsec:recovering_session_parameters}),
blind-decodes the corresponding PDCCH candidates,
and tags each recovered DCI with its RNTI, CORESET,
and search space.

\subsubsection{DCI Interpretation and Validation}
\label{subsubsec:false_positive_dci}

\sys first classifies each DCI by its RNTI type
(SI-, P-, RA-, TC-, or C-RNTI) and uses the DCI
format-indicator bit to distinguish the uplink and
downlink formats where applicable~\cite{3gpp38212}.
DCIs with SI-, P-, RA-, or TC-RNTIs use cell-common
configurations, whereas C-RNTI DCIs may require
UE-dedicated configurations.

Fallback formats 0\_0 and 1\_0 have BWP-dependent
layouts, although their field interpretation may
depend on UE-dedicated configurations.
Non-fallback formats 0\_1 and 1\_1 additionally
have configuration-dependent field widths
(\autoref{tab:dci11_setup_vs_reconfig}).
For each matching-size profile, \sys derives the
field layout, unpacks the DCI, and interprets the
resource allocation and MCS using the profile's
configuration~\cite{3gpp38212,3gpp38214}.
This produces one scheduling-grant hypothesis
per candidate profile.

To eliminate false positives, \sys checks the DCI
CRC and the consistency of the resulting grant.
Only the last 16 bits of the 24-bit DCI CRC are
RNTI-scrambled~\cite{3gpp38212}. Thus, XORing the
received CRC with the CRC recomputed over the DCI
payload must yield zero in the first 8 bits.
\sys also checks that the recovered RNTI matches
the hypothesis or is valid for the search space,
and rejects invalid field interpretations, such as
resource blocks outside the BWP, nonexistent allocation
entries, or reserved codepoints.
The transport-block CRC subsequently confirms
the correct profile
(\autoref{subsubsec:pdsch_pusch_decoding}).

\subsubsection{PDSCH and PUSCH Decoding}
\label{subsubsec:pdsch_pusch_decoding}

Each scheduling grant specifies the transmission
resources, MCS, layers, and DMRS ports, while the
matching profile supplies the remaining parameters.
\sys applies the standard receiver chain
(\autoref{fig:decoding_chains})
~\cite{3gpp38211,3gpp38212,3gpp38214}.

\nibf{Grant verification.}
\sys decodes the grants derived from candidate
profiles and uses the transport-block CRC to
identify the correct one. It caches the verified
profile and PDSCH DMRS scrambling identity,
obtained from RRC Setup or DMRS correlation
(\autoref{subsubsec:dmrs_id_recovery}),
and prioritizes them for subsequent transmissions.

\nibf{Scheduling and resource mapping.}
Downlink grants typically schedule PDSCH in the
current slot ($k_0=0$), whereas uplink grants
schedule PUSCH at offset $k_2$.
\sys stores uplink grants by their target slots
and processes them when those slots arrive.
Grants within a slot share one FFT per antenna.
For PDSCH, resource mapping excludes resource elements
occupied by DMRS code-division multiplexing (CDM) groups without data, non-zero-power (NZP) or zero-power (ZP) CSI-RS,
SSBs, and PDCCH according to the profile. For PUSCH,
\sys applies the resource allocation and DMRS configuration
specified by the uplink grant and profile.

\nibf{MIMO decoding.}
For DCI format 1\_1, \sys obtains the number of
layers and DMRS ports from the antenna-port field
using the profile~\cite{3gpp38212}.
Using the indicated DMRS ports, \sys estimates the
effective channel for each layer and separates the
layers with minimum mean-square error (MMSE) equalization, without requiring
the precoding matrix or CSI reports.
\sys supports up to $4\times4$ downlink MIMO.

\subsubsection{Adaptive Uplink Synchronization}
\label{subsubsec:adaptive_ul_sync}

Downlink-derived slot timing cannot directly
synchronize uplink reception: each UE applies
timing advance relative to the gNB, so its
signal reaches the sniffer with a different offset.
As observed by LTESniffer~\cite{hoang:ltesniffer},
a timing offset within the cyclic prefix introduces
a subcarrier-dependent phase rotation that channel
estimation can compensate.
Thus, \sys need not align exactly with each UE;
it only needs to keep all UE arrivals within the
receiver's tolerable timing window.

\sys estimates each UE's arrival-time offset from
the DMRS of its scheduled PUSCH and gradually
adjusts the synchronization window toward the
center of the observed arrival-time range.
To prevent false DCIs or
incorrect profiles from corrupting synchronization,
only repeated, consistent DMRS detections contribute
to the adjustment. \sys continuously updates
the window as UEs connect, disconnect, or move.

\subsubsection{Active-UE Management}
\label{subsubsec:active_ue_management}

\sys maintains an active-UE table keyed by C-RNTI,
storing each UE's verified profile, session-specific
parameters, and last activity time.
For newly connecting UEs, \sys initializes a UE context
using the TC-RNTI observed in the RAR and updates the
table key to the corresponding C-RNTI after successful
contention resolution. For already-connected UEs,
\sys creates an entry after recovering a C-RNTI and
successfully interpreting a DCI
(\autoref{subsubsec:syndrome_rnti_recovery}).
Each valid DCI refreshes the UE's activity time,
and active RNTIs receive priority during PDCCH
decoding. 
Entries inactive for 10 seconds are removed to bound
stale state and reduce the risk of reusing outdated
configuration or session context after RNTI reassignment.

\subsection{Real-Time CPU--GPU Decoding}
\label{subsec:design_realtime}

At 30\,kHz subcarrier spacing, each slot lasts
0.5\,ms, so real-time operation requires \sys to
sustain a processing throughput of 2,000 slots per second.
IQ transfer and OFDM processing scale with bandwidth and
antenna count; for example, capturing a 100\,MHz cell
with four antennas requires two 10\,GbE links
(\autoref{subsec:eval_setup}).
Unlike a UE, which monitors only its own PDCCH candidates,
\sys searches across active RNTIs, recovers parameters of
previously unidentified UEs, and decodes their PDSCH and
PUSCH transmissions. Thus, its decoding workload grows
with the number of candidates and scheduled transmissions.

To handle this workload, \sys combines parallel CPU
Slot Workers with GPU-accelerated candidate searches.
The GPU performs batched PDCCH/DCI decoding,
syndrome-guided C-RNTI recovery
(\autoref{subsubsec:syndrome_rnti_recovery}), and
PDSCH DMRS scrambling-identity correlation
(\autoref{subsubsec:dmrs_id_recovery}).
These operations evaluate many independent hypotheses
and are well suited to GPU parallelism.
Each CPU Slot Worker processes one slot, allowing
multiple slots to be processed concurrently.
When all workers are busy, \sys drops newly arriving
slots instead of queuing them. This prevents an
accumulating backlog and bounds processing latency,
at the cost of reduced decoding coverage under
heavy load.

For each slot, a worker computes PDCCH LLRs and prepares CCE locations, candidate DCI
sizes, and active RNTIs.
The GPU evaluates the candidates and returns CRC-valid
DCIs, which the worker interprets into PDSCH and PUSCH
grants (\autoref{subsec:configuration_aware_end_to_end_decoding}).
For PDSCH grants with unknown DMRS scrambling identities,
the GPU performs correlation to recover the identities.
The worker then decodes the scheduled transport blocks.

\subsection{Implementation}
\label{subsec:implementation}

We implement \sys on top of OAI
~\cite{openairinterface5g}, adding approximately 53k
lines of C, C++, and CUDA code and modifying 8k lines
of existing code. The added code implements cell search,
configuration-profile discovery and management,
multi-UE PDCCH/DCI processing, session-parameter recovery,
PDSCH/PUSCH decoding, adaptive uplink synchronization,
active-UE management, and parallel Slot Workers.

We adapt the OAI UE's PHY and medium access control (MAC) procedures to operate
as a passive multi-UE receiver. Instead of following the
connection state of a single UE, the modified procedures
use recovered DCIs and configuration profiles to decode
the uplink and downlink traffic of multiple UEs. 
We adapt OAI's channel decoders and radio drivers for
passive sniffing and validate the receiver chain against
a floating-point reference decoder built with
Sionna~\cite{sionna}.

\nibf{GPU acceleration.}
We adapt the NVIDIA Aerial cuPHY polar decoder~\cite{cuphy}
to batch-decode PDCCH candidates with different DCI sizes
and RNTI hypotheses.
We also implement CUDA kernels for parallel
frozen-bit syndrome evaluation and PDSCH DMRS
scrambling-identity correlation.
Our evaluation uses an NVIDIA RTX 3080~Ti.

\nibf{Hardware and I/O.}
\sys supports uplink and downlink reception through
UHD-compatible USRP radios, including the B200, B210,
X310, and N310.
The supported bandwidth and MIMO configuration depend
on the radio: the B200 supports SISO up to 40\,MHz,
the X310 supports $2\times2$ MIMO up to 100\,MHz,
and the N310 supports $4\times4$ MIMO up to 100\,MHz.
\sys also supports IQ recording and replay and exports
decoded MAC protocol data units (PDUs) as PCAP files for Wireshark analysis.

\section{Evaluation}
\label{sec:evaluation}
We evaluate \sys on a controlled testbed and two
commercial 5G SA networks in different countries
to answer five research questions:
\nibf{RQ1:} Can \sys correctly decode over-the-air 5G NR downlink and uplink traffic?~(\autoref{subsec:eval_accuracy})

\nibf{RQ2:} Can capability-guided profiling discover PHY configurations assigned by commercial networks after encrypted RRC Reconfiguration?~(\autoref{subsec:eval_profiling})

\nibf{RQ3:} How much do configuration profiles and identifier recovery contribute to decoding?~(\autoref{subsec:eval_ablation})

\nibf{RQ4:} Under what network loads can \sys sustain real-time decoding of wideband commercial traffic?~(\autoref{subsec:eval_realtime})
 
\nibf{RQ5:} How does \sys compare with existing 5G NR sniffers?~(\autoref{subsec:eval_comparison})

\subsection{Experimental Setup}
\label{subsec:eval_setup}

\nibf{Controlled testbed.}
We deploy an srsRAN gNB~\cite{srsran5g} on a 40\,MHz n41 channel with three commercial smartphones: Xiaomi 15 Ultra, Google Pixel 8 Pro, and Redmi Note 9T.
The gNB, \sys, and SNI5GECT use USRP X310 radios and PCs with an Intel i7-14700K CPU and 32\,GB RAM; \sys runs without a GPU due to low load.
We evaluate downlink SISO and 2$\times$2 MIMO with the two sniffers colocated in three placements: S1 (gNB, UEs, and sniffers colocated), S2 (gNB and sniffers together, 18\,m from UEs), and S3 (an 18\,m-side triangle).
In each run, we start the sniffers, disable airplane mode on the UEs, and visit a website.
We repeat each setting three times.

\nibf{Commercial networks.}
We evaluate two 100\,MHz, 30\,kHz SCS 5G SA networks: Network~K (Country~K, n78; SSB SINR 39\,dB) and Network~U (Country~U, n41; 22\,dB).
\sys uses a four-antenna USRP N310 connected via two 10\,GbE links to an Intel i7-14700K PC with 32\,GB RAM and an NVIDIA RTX 3080~Ti GPU.
The links carry the four-channel 100\,MHz IQ stream.
For target-UE experiments, we use our own Samsung Galaxy Z Flip 4 on Network~K and Xiaomi Redmi Turbo 4 on Network~U, identifying their traffic by XCAL-reported C-RNTIs. 

\nibf{OAI UE-based configuration discovery.}
We run the modified OAI UE~(\autoref{subsubsec:ue_config_discovery}) on a USRP N310 and an Intel i7-14700K PC with 32\,GB RAM.
For each replayed UE capability, the OAI UE registers using a commercial SIM in a PC/SC reader and logs the decrypted RRC Setup and RRC Reconfiguration messages.

\nibf{Metrics.}
On the testbed, we use srsRAN gNB logs as ground truth to evaluate decoded DCIs and PDSCH/PUSCH transport-block coverage.
Without commercial-network ground truth, we compare
CRC-valid DCI and PDSCH/PUSCH transport-block counts
with and without each technique.
For target UEs, we also report the decoding success rate: CRC-valid PDSCH/PUSCH transport blocks divided by decoded DL/UL DCIs, respectively.
This conditional rate does not measure coverage of all transmitted traffic.

\subsection{RQ1: Decoding Correctness and Coverage}
\label{subsec:eval_accuracy}
\begin{figure*}[t]
\centering
\includegraphics[width=.9\textwidth]{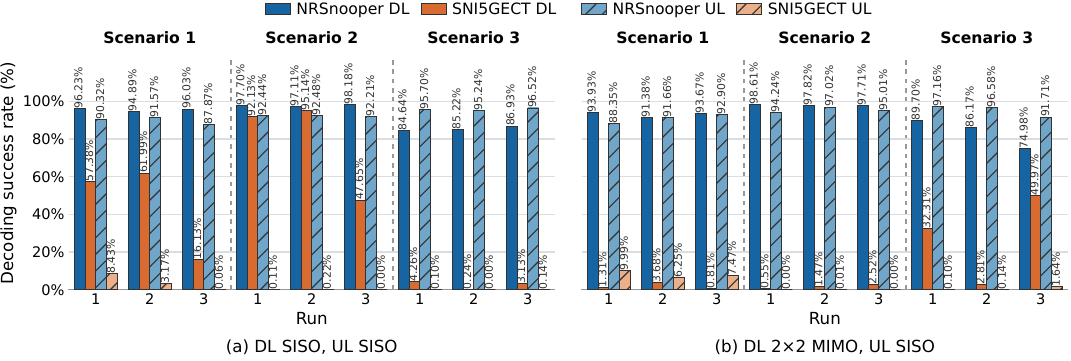}
\caption{Testbed transport-block decoding coverage of
\sys and SNI5GECT: (a) downlink SISO and
(b) downlink 2$\times$2 MIMO, across three placements
and runs. Solid/hatched bars indicate downlink/uplink.}
\label{fig:eval_testbed}
\end{figure*}

We evaluate correctness and coverage against the srsRAN gNB logs.
All CRC-valid transport blocks decoded by \sys match the gNB payloads byte-for-byte.
Across all placements and downlink SISO/2$\times$2 MIMO configurations, \sys recovers 74.98--98.61\% of downlink and 87.87--97.16\% of uplink transport blocks~(\autoref{fig:eval_testbed}).
Downlink coverage is lowest in S3, where the sniffer is 18\,m from the gNB.
This is consistent with differences between the UE and sniffer channels: the gNB selects MCS, rank, and precoding based on the UE's CSI.
The mismatch can be more pronounced under MIMO, where the sniffer separates layers using precoding selected for the UE's channel.
Thus, \sys decodes transport blocks correctly, with coverage depending on receiver placement and channel conditions.

\subsection{RQ2: Configuration Profiling}
\label{subsec:eval_profiling}

\nibf{Discovered profiles.}
We replay 104 UE capabilities on two commercial networks: 21 collected from COTS UEs spanning five phone and five baseband vendors~(\autoref{tab:app_cots}), and 83 variants obtained by mutating 57 fields and several band combinations~(\autoref{subsubsec:capability_mutation}).
We discover 21 profiles on Network~K and 18 on Network~U~(\autoref{tab:eval_profiles}; full lists in~\autoref{tab:app_profiles_k} and~\autoref{tab:app_profiles_u}).
Only 11 profiles (5 on K and 6 on U) arise from the original capabilities; mutations reveal the remaining 28.

The networks apply different configuration policies.
Network~K adapts UL precoding to UE MIMO capabilities, whereas Network~U uses the same UL configuration across the tested capabilities.
Of the 57 mutated fields, only 20, plus band combinations, affect RRC Reconfiguration, and only 14 select distinct profiles~(\autoref{tab:app_cap_fields}).
Both networks select MCS tables based on 256QAM support and expand the DCI 1\_1 resource-allocation field from 16 to 19 bits for dynamic allocation (K12, U03).

These differences affect decoding beyond DCI size.
DCI 1\_1 sizes range from 50 to 57 bits, yet equal-size profiles may have different field layouts (K05/K06) or MCS tables (K16/K01), affecting scheduling interpretation and transport-block decoding.
Unlike LTE, where only a few UE-specific decoding parameters need testing~\cite{hoang:ltesniffer}, NR requires configuration-aware profiling rather than blind parameter combinations.

\begin{table}[t]
\centering
\footnotesize
\setlength{\tabcolsep}{4pt}
\caption{Replayed UE capabilities and discovered profiles.}
\label{tab:eval_profiles}
\begin{tabular}{lr}
\toprule
COTS UE capabilities & 21 \\
Mutated variants & 83 \\
Fields mutated & 57 + Band Combinations \\
\midrule
Distinct profiles, Network~K & 21 \\
Distinct profiles, Network~U & 18 \\
\bottomrule
\end{tabular}
\end{table}

\nibf{Validation against commercial UEs.}
We validate the learned profiles against our Galaxy Z Flip 4 on Network~K and Redmi Turbo 4 on Network~U.
For each phone, XCAL records its C-RNTI and RRC Reconfiguration as it reconnects and accesses a website.
We compare these configurations with those obtained by replaying the same phone's capabilities on the OAI UE~(\autoref{tab:app_cots}).
Both phones receive matching UE-dedicated PHY configurations (K02 in~\autoref{tab:app_profiles_k} and U06 in~\autoref{tab:app_profiles_u}).
Using these profiles, \sys decodes 86.7\%/96.1\% of PDSCH/PUSCH transport blocks scheduled by decoded DCIs on Network~K, and 61.0\%/85.1\% on Network~U~(\autoref{fig:eval_commercial}(c)--(d)).
We examine cell-wide gains in~\autoref{subsec:eval_ablation}.

\subsection{RQ3: Contribution of Each Technique}
\label{subsec:eval_ablation}

\begin{figure*}[t]
\centering
\includegraphics[width=.9\textwidth]{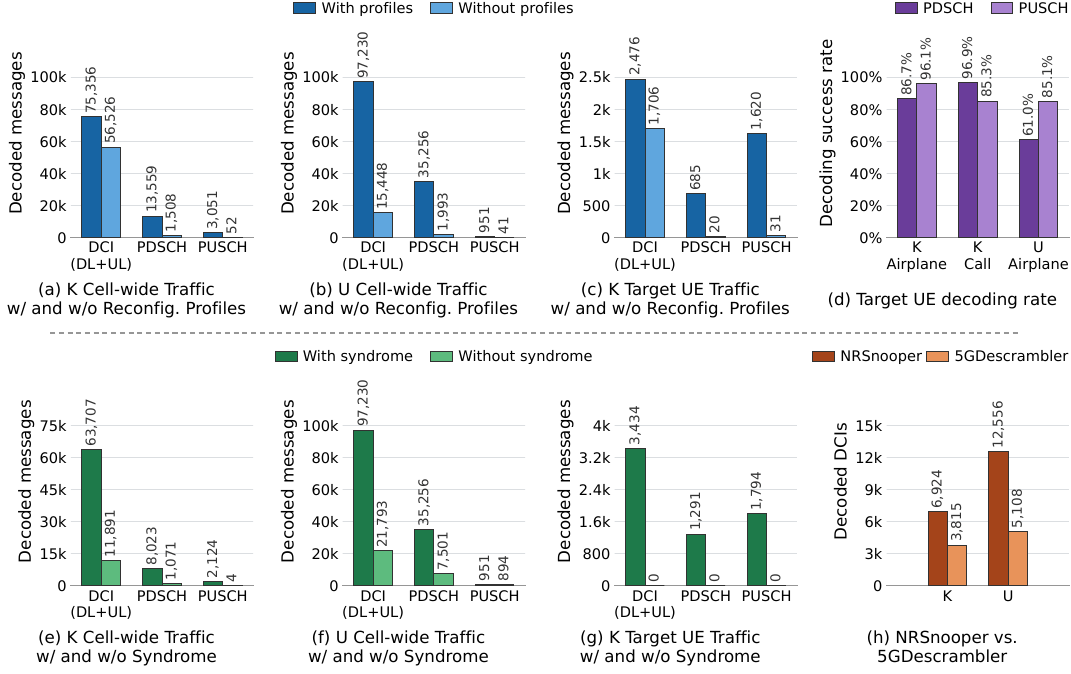}
\caption{Commercial-network results: (a)--(c) profile ablation; (d) target-UE decoding rate; (e)--(g) RNTI-recovery ablation; (h) DCI comparison with 5GDescrambler.}
\label{fig:eval_commercial}
\end{figure*}

\nibf{Configuration profiles.}
We evaluate cell-wide decoding on 60-second captures from each commercial network, with and without the OAI-derived profiles.
On Network~K, profiles increase decoded DCIs, PDSCH, and PUSCH transport blocks by 1.3$\times$, 9.0$\times$, and 58.7$\times$, respectively; on Network~U, by 6.3$\times$, 17.7$\times$, and 23.2$\times$~(\autoref{fig:eval_commercial}(a)--(b)).
For our own UE on Network~K, profiles increase decoded PDSCH and PUSCH transport blocks from 20 to 685 and from 31 to 1,620~(\autoref{fig:eval_commercial}(c)).
The smaller DCI gain on Network~K occurs because many DCIs retain their RRC Setup sizes and remain decodable without profiles.
However, changes to PDSCH/PUSCH configurations (\eg, MCS tables) prevent correct grant interpretation and data decoding.
Some discovered profiles match no traffic in these captures because the corresponding UEs are inactive; \sys retains them for future connections.
These results demonstrate the importance of configuration profiles for decoding traffic after encrypted RRC Reconfiguration.

\nibf{Identifier recovery.}
We evaluate syndrome-guided RNTI recovery for already-connected UEs.
On Network~K, we start \sys during an ongoing phone call, after our UE has connected.
Without RNTI recovery, \sys decodes none of its traffic; with recovery, it decodes 3,434 DCIs, 1,291 PDSCH, and 1,794 PUSCH transport blocks~(\autoref{fig:eval_commercial}(g)).

For cell-wide traffic~(\autoref{fig:eval_commercial}(e)--(f)), RNTI recovery increases decoded DCIs and PDSCH transport blocks by 5.4$\times$ and 7.5$\times$ on Network~K, while PUSCH transport blocks increase from 4 to 2,124.
On Network~U, DCI and PDSCH decoding improve by 4.5$\times$ and 4.7$\times$, but PUSCH decoding changes little (894 to 951).
Most uplink traffic comes from our test smartphone, which connects after sniffing begins, allowing \sys to obtain its C-RNTI from the RAR without syndrome recovery.
Thus, RNTI recovery is essential for decoding UEs whose initial access was not observed.

\subsection{RQ4: Real-Time Decoding}
\label{subsec:eval_realtime}

\begin{figure*}[t]
\centering
\includegraphics[width=.9\textwidth]{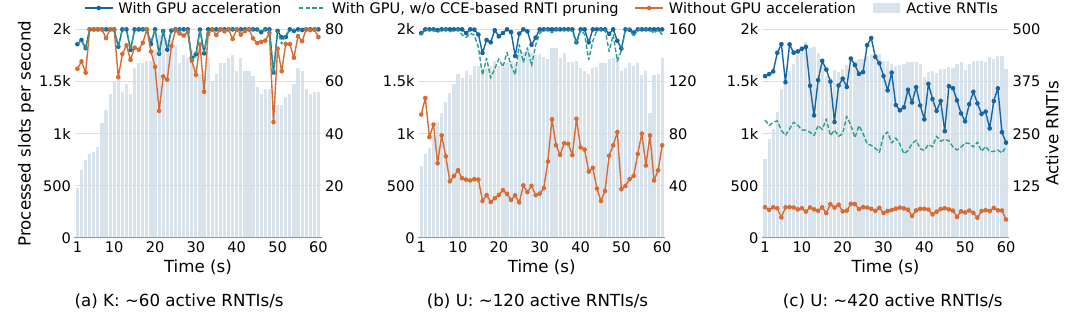}
\caption{Processed slots/s under increasing loads:
(a) Network~K (60 active RNTIs/s),
(b) Network~U (120), and (c) Network~U (420).
Curves compare GPU, GPU without CCE pruning, and CPU-only decoding.
Bars show active RNTIs/s; the real-time target is 2,000 slots/s.}
\label{fig:eval_realtime}
\end{figure*}

At 30\,kHz subcarrier spacing, real-time decoding requires processing 2,000 slots per second.
Each slot carries 100\,MHz of samples from four receive antennas, while decoding workload increases with active UEs and scheduled transmissions~(\autoref{subsec:design_realtime}).
We capture 60 seconds of traffic on Network~K, Network~U in the morning, and Network~U at peak hour, with approximately 60, 120, and 420 active RNTIs per second, respectively.
We replay each capture at 2,000 slots per second using the live decoding pipeline.
When all Slot Workers are busy, \sys drops incoming slots; \autoref{fig:eval_realtime} reports the processed slots.

On Network~K, GPU processing stays near 2,000 slots per second, while CPU-only processing occasionally falls to 1,100.
On Network~U with 120 active RNTIs per second, the GPU processes 1,750--2,000 slots per second, compared with 350--1,350 without it.
Under heavy load (420 active RNTIs per second), throughput falls to 900--1,900 slots per second with the GPU and 170--320 without it.

Thus, GPU acceleration enables near-real-time bidirectional decoding of 100\,MHz commercial traffic under moderate load, including Post-Reconf and already-connected UE traffic, with occasional dropped slots.
Heavy load exceeds the capacity of the tested GPU, requiring further optimization or greater processing capacity.

\nibf{Effect of CCE-based RNTI pruning.}
We evaluate A2.2 by disabling CCE-based pruning, forcing the GPU to test all $2^{15}$ RNTI signatures per PDCCH candidate~(\autoref{fig:eval_realtime}).
Under light load (Network~K, 60 RNTIs/s), both configurations process 97.7\% of slots.
Under moderate load (Network~U, 120 RNTIs/s), it increases processed slots from 94.6\% to 98.4\%.
Under heavy load (Network~U, 420 RNTIs/s), the gain grows from 47.7\% to 72.5\%.

\subsection{RQ5: Comparison with Existing Sniffers}
\label{subsec:eval_comparison}

\nibf{SNI5GECT.}
We compare \sys with SNI5GECT~\cite{luo2025sni5gect}
on identical testbed captures~(\autoref{fig:eval_testbed}).
The testbed provides favorable conditions for SNI5GECT:
RRC Reconfiguration is unencrypted, captures begin before
UE connections, and traffic load is low.
Thus, C1--C3~(\autoref{subsec:challenges_approaches})
are not exercised, allowing us to compare decoding capabilities.

Even under these conditions, \sys recovers
74.98--98.61\% of downlink and 87.87--97.16\% of uplink
transport blocks, versus 0.24--95.14\% and 0--9.99\%
for SNI5GECT.
In SISO, \sys achieves higher downlink coverage, including
for 64QAM and 256QAM transmissions.
SNI5GECT does not support downlink MIMO, explaining
its poor coverage under 2$\times$2 MIMO.
Its uplink decoding is further limited by unsupported
DCI 0\_1 configurations with 2-port SRS and an IQ
buffer-copying bug, preventing a direct comparison
of PUSCH decoder quality.

These results expose the limited decoding coverage of
SNI5GECT even in a favorable testbed.
More importantly, unlike SNI5GECT, \sys decodes
Post-Reconf traffic and already-connected UEs on
commercial networks, as demonstrated in
\autoref{subsec:eval_ablation}.

\nibf{5GDescrambler.}
5GDescrambler~\cite{windisch2026descrambler}
recovers DCIs without knowing the RNTI or PDCCH
DMRS scrambling identity, but does not decode PDSCH
or PUSCH.
We compare both decoders using identical PDCCH
configurations and LLRs extracted from 10-second
commercial-network captures, isolating their DCI
decoding capabilities.

On Network~K, \sys decodes 6,924 DCIs versus 3,815
for 5GDescrambler (1.8$\times$); on Network~U,
12,556 versus 5,108 (2.5$\times$)
~(\autoref{fig:eval_commercial}(h)).
5GDescrambler lacks full polar error correction,
whereas \sys combines syndrome-guided RNTI ranking
with full soft-decision polar decoding and CRC
validation~(\autoref{subsec:recovering_session_parameters}).

Thus, even with identical PDCCH inputs from commercial
networks, \sys recovers substantially more DCIs.
Unlike 5GDescrambler, \sys also interprets these DCIs
using recovered configurations and decodes the
corresponding PDSCH and PUSCH traffic.
\section{Discussion and Limitations}
\label{sec:discussion_limitations}

\nibf{Per-channel PHY measurements.}
Besides decoded traffic, \sys obtains signal-to-noise ratio (SNR) measurements
and channel estimates from the DMRS of each decoded
PDCCH, PDSCH, and PUSCH transmission.
Associated with individual RNTIs, these measurements
can support tracking changes in observed radio conditions
over time. However, they reflect the channel at the
sniffer rather than at the intended receiver; in particular,
downlink measurements do not represent the channel
experienced by the target UE.
We leave such applications to future work.

\nibf{Narrowing down candidate phone models.}
A decoded UE's configuration profile can narrow its
candidate phone models using the capability-to-profile
mapping discovered during profiling 
(\autoref{tab:app_profiles_k} and~\autoref{tab:app_profiles_u}).
However, multiple models may share a profile:
K14 covers five Xiaomi models, while K02 includes
Samsung, Apple, and Xiaomi devices.
Combining profiles with non-access stratum (NAS) security capabilities
~\cite{oh2023preventing}, timing advance
~\cite{kotuliak2022ltrack}, and MAC control elements
could improve fingerprinting.
We have not evaluated such identification and leave
it to future work.

\nibf{Potential Countermeasures.}
The confidentiality risks of lower-layer cellular
signaling have been recognized since LTE, including 
in LTESniffer and UMA~\cite{oh2024enabling}.
However, retrofitting fundamental confidentiality
protection into deployed 5G NR faces substantial
compatibility and performance constraints.
We therefore argue that 6G should establish explicit
requirements for lower-layer confidentiality.

In particular, 6G should consider protecting sensitive
UE-specific scheduling metadata, radio identifiers,
and information enabling observers to reconstruct
or link a UE's uplink and downlink activities.
Such protection should complement end-to-end
encryption while preserving fast initial access,
efficient UE-specific control-channel reception
without decoding other UEs' messages, low scheduling
latency, limited control-channel bandwidth, and
acceptable computational, signaling, and energy overhead.

Encrypting UE capabilities or RRC Reconfiguration
alone is insufficient, as \sys learns reusable PHY
configurations through legitimate connections.
Session-specific scrambling identities and randomized
PHY configurations may increase decoding costs but
do not guarantee confidentiality.
Designing protection mechanisms that satisfy both
security and efficiency requirements remains an
open research problem for 6G.

\nibf{Limitations.}
First, although sniffing is passive, configuration
profiling requires legitimate network access through
an OAI UE with a commercial SIM.
Some networks also require an accepted IMEI.
Profiles remain reusable only while
the relevant network configurations remain unchanged.

Second, profile coverage is incomplete: a gNB may
assign configurations not triggered by our collected
or mutated capabilities. Broader capability collection
and testing across more gNBs could improve coverage.

Third, decoding depends on signal quality at the
sniffer. The gNB adapts MCS, rank, and precoding
to the intended UE's channel, which may differ
from the sniffer's~(\autoref{subsec:eval_accuracy}).
Beamforming can further reduce reception outside
the intended beam. Closer placement or multiple
sniffers could mitigate this limitation.

Fourth, real-time wideband, multi-antenna decoding
requires substantial SDR I/O and processing capacity.
Our 100\,MHz implementation uses a USRP N310 and
GPU acceleration, yet cannot sustain real-time
processing under the heaviest observed load~(\autoref{subsec:eval_realtime}).
Narrower-bandwidth or selective monitoring could help but remains unevaluated.
\section{Related Work}
\label{sec:related_work}

\nibf{LTE sniffers.}
LTEye~\cite{kumar2014lte}, OWL~\cite{bui2016owl},
FALCON~\cite{falkenberg2019falcon}, and
NG-Scope~\cite{xie2022ngscope} recover LTE scheduling
information for network measurement.
LTESniffer~\cite{hoang:ltesniffer} further decodes
bidirectional data traffic in real time, exploiting LTE's
limited UE-specific PHY configurations.
In contrast, MobileInsight~\cite{li2016mobileinsight},
QXDM~\cite{qualcomm_qxdm}, and XCAL~\cite{xcal}
observe only their host devices.
These approaches do not address NR's extensive
UE-dedicated configurations delivered through encrypted
RRC signaling~(\autoref{sec:motivation}).

\nibf{5G sniffers.}
\autoref{tab:prior_work} compares the channel,
connection-state, and real-time coverage of existing
research sniffers.
5G-Sniffer~\cite{ludant20235g} searches over unknown
PDCCH parameters, whereas
5GDescrambler~\cite{windisch2026descrambler} exploits
the algebraic structure of scrambling to recover DCI
without knowing these parameters.
Both recover control information but do not provide
general PDSCH/PUSCH decoding.

NR-Scope~\cite{wan2024nrscope} decodes PDCCH and
PDSCH using configurations acquired during initial
access, while SNI5GECT~\cite{luo2025sni5gect}
supports real-time bidirectional sniffing during
connection establishment to enable message injection.
Both depend on observing initial signaling and do not
recover encrypted configuration updates or missing
session parameters of already-connected UEs.

Golden Sniffer~\cite{palama2026golden} exploits Gold-sequence structure to recover PDCCH DMRS scrambling identities and RNTIs, and decodes PDSCH scheduled by fallback DCI.
However, it does not demonstrate non-fallback DCI interpretation or PUSCH decoding in the Post-Reconf state in commercial networks.
\sys instead discovers reusable Post-Reconf PHY profiles and recovers session-specific parameters to decode already-connected UEs bidirectionally.

Commercial analyzers such as
WaveJudge~\cite{wavejudge} also provide real-time
5G NR decoding. However, public documentation does
not establish whether they can recover unknown
UE-specific configurations and session parameters
of arbitrary already-connected UEs under our threat model.

\nibf{Syndrome-guided RNTI recovery.}
Gardner and Roth~\cite{gardner2020efficient} proposed
frozen-bit syndrome matching for RNTI recovery.
Richards et al.~\cite{richards2024rnti} reduced the
search to $2^{15}$ scrambling classes and confirmed
candidates through polar decoding and CRC validation.
Both evaluated their approaches in simulation.
Building on these techniques, \sys exploits the
CCE--RNTI search-space relationship to prune candidate
hypotheses, reducing tested candidates by 22\% and
81\% in Networks~K and~U, respectively~(\autoref{subsubsec:rnti_candidate_reduction}).
It integrates syndrome-based ranking, soft-decision
polar decoding, and CRC validation into a real-time
commercial NR decoder, using recovered RNTIs for
subsequent data decoding.

\nibf{Applications of sniffers.}
Cellular sniffing enables identity mapping, traffic
fingerprinting, and user localization
~\cite{rupprecht2019breaking,bae2022,
kotuliak2022ltrack,oh2024enabling}.
It also supports fake-base-station detection
~\cite{ney2017seaglass,tucker2025marlin,oh2026devilray}.
In 5G, 5G-Sniffer~\cite{ludant20235g} demonstrates
phone-number-to-RNTI linkage, while
SNI5GECT~\cite{luo2025sni5gect} uses sniffed traffic
to time message injection.
\sys provides the bidirectional visibility needed
to revisit such analyses in ongoing commercial
5G SA connections.

\nibf{UE capability and device fingerprinting.}
Previous works exploit UE capabilities for
fingerprinting, bidding-down attacks, and SIM-box
detection~\cite{shaik2019new,karakoc2023never,
oh2023preventing}.
Instead, \sys replays capabilities via
its own UE to discover the PHY configurations assigned
by commercial gNBs, without identifying a sniffed UE's exact model.
Matching profiles may nevertheless narrow possible
UE models~(\autoref{sec:discussion_limitations}).

\nibf{Lower-layer attacks.}
Unprotected lower-layer information and insufficient
integrity protection enable traffic analysis,
packet manipulation, and impersonation
~\cite{kohls2019lost,rupprecht2019breaking,
rupprecht2020imp4gt}.
PHY-layer attacks such as SigOver~\cite{yang2019hiding},
AdaptOver~\cite{erni2022adaptover}, and
SigUnder~\cite{ludant2021sigunder} manipulate radio
transmissions. Such attacks can benefit from the
timing, scheduling, and PHY parameters revealed
by passive sniffing.

\nibf{Testing lower layers.}
Prior work actively tests cellular systems through
malformed signaling~\cite{kim2019touching,
park2022doltest,hussain2021noncompliance},
lower-layer fuzzing~\cite{hoang2025llfuzz,
matheus2023_5ghoul}, baseband
emulation~\cite{maier2020basesafe,
hernandez2022firmwire}, and specification-based
analysis~\cite{kim2021basespec,kim2023basecomp}.
In contrast, \sys passively observes commercial networks and could support future testing deployed PHY configurations.

\section{Conclusion}

We presented \sys, a system for real-time bidirectional
decoding of commercial 5G NR traffic. By combining
capability-guided configuration profiling, session-specific
parameter recovery, and a hybrid CPU--GPU pipeline,
\sys decodes traffic after encrypted RRC Reconfiguration
and from already-connected UEs without access to their
cryptographic keys. Evaluations on two commercial 5G SA
networks demonstrate substantial improvements in decoded
traffic and bidirectional decoding during an ongoing
phone call.

Beyond security and privacy research, \sys could support
network troubleshooting, performance analysis, and
security monitoring. However, its ability to link uplink
and downlink activities to individual UEs also raises
privacy concerns. Improving scalability under heavy
traffic and extending deployment coverage remain future
work. More fundamentally, our findings motivate explicit
lower-layer confidentiality requirements for 6G while
preserving the efficiency and performance of cellular
communication.
\section*{Ethics Considerations}
\label{sec:ethics}

\nibf{Controlled testbed.} We conducted all testbed experiments, including the collection of UE capability information from commercial smartphones, on an open-source 5G network using a test PLMN identity and test SIM cards. The testbed operated inside an isolated room, so its signals did not interfere with commercial networks, and commercial UEs outside the room could not attach to it.

\nibf{Commercial networks.} \sys is a passive sniffer that never transmits; it decodes the unencrypted downlink control information and the uplink and downlink transport blocks, without decrypting their encrypted payloads. Only the configuration-discovery phase transmits: our instrumented open-source UE follows the standard registration and authentication procedures using a legitimate commercial SIM from our own subscriptions and records the RRC configurations assigned to its own connection. The UE reported capability information conforming to TS 38.331, including mutated variants of real smartphone capabilities, and the small number of connections we made placed negligible load on the network.
When decoding commercial traffic, we only verified whether each control message and data block passed its CRC check.
For the ablation and real-time experiments, which replay 60-second captures, our scripts stored the raw IQ samples only temporarily during each run and deleted them automatically afterward.
We retained only terminal and system logs with aggregate statistics, such as CRC outcomes and the number of decoded messages and data blocks.
We did not store any decoded packets, did not attempt to decrypt protected traffic, and did not attempt to link radio identifiers to subscribers.

\nibf{Targeted UE experiment.} For the experiment with an already-connected UE, we used only phones and subscriptions owned by the authors. We obtained our UE's radio identifier from a commercial diagnostic tool (XCAL~\cite{xcal}) and used it to isolate our own traffic, discarding traffic from all other UEs.





\bibliographystyle{IEEEtran}
\bibliography{p,sslab,conf}


\appendices

\section{DCI Format 1\_1 Fields}
\label{sec:app_dci_fields}

\autoref{tab:dci11_fields_appendix} lists all fields of DCI format 1\_1 and the RRC parameters that set the width of each field.

\begin{table}[t]
\centering
\begin{threeparttable}
\caption{DCI format 1\_1 fields in TS~38.212 Rel-15~\cite{3gpp38212} and the RRC parameters in TS~38.331~\cite{3gpp-5g-rrc} that determine their sizes.}
\label{tab:dci11_fields_appendix}
\scriptsize
\setlength{\tabcolsep}{3pt}
\renewcommand{\arraystretch}{1.08}
\begin{tabular}{@{}c >{\raggedright\arraybackslash}p{84pt} >{\centering\arraybackslash}p{20pt}
  >{\raggedright\arraybackslash\hyphenpenalty=10000}p{118pt}@{}}
\toprule
\textbf{No.} & \textbf{Field} & \textbf{Size (bits)} & \textbf{Configuration (Channel $\rightarrow$ parameter)} \\
\midrule
1  & Identifier for DCI formats               & 1       & -- \\
2  & Carrier indicator                        & 0, 3    & Cell $\rightarrow$ cif-Presence \\
3  & Bandwidth part indicator                 & 0--2    & Cell $\rightarrow$ downlinkBWP-ToAddModList \\
4  & Frequency domain resource assignment     & Var.    & PDSCH $\rightarrow$ resourceAllocation, rbg-Size\newline BWP $\rightarrow$ locationAndBandwidth \\
5  & Time domain resource assignment          & 0--4    & PDSCH $\rightarrow$ pdsch-TimeDomainAllocationList \\
6  & VRB-to-PRB mapping                       & 0, 1    & PDSCH $\rightarrow$ vrb-ToPRB-Interleaver \\
7  & PRB bundling size indicator              & 0, 1    & PDSCH $\rightarrow$ prb-BundlingType \\
8  & Rate matching indicator                  & 0--2    & PDSCH $\rightarrow$ rateMatchPatternGroup1/2 \\
9  & ZP CSI-RS trigger                        & 0--2    & PDSCH $\rightarrow$ aperiodic-ZP-CSI-RS-ResourceSets\-ToAddModList \\
10 & Modulation and coding scheme (TB~1)      & 5       & -- \\
11 & New data indicator (TB~1)                & 1       & -- \\
12 & Redundancy version (TB~1)                & 2       & -- \\
13 & Modulation and coding scheme (TB~2)      & 0, 5    & PDSCH $\rightarrow$ maxNrofCodeWordsScheduledByDCI \\
14 & New data indicator (TB~2)                & 0, 1    & As No.~13 \\
15 & Redundancy version (TB~2)                & 0, 2    & As No.~13 \\
16 & HARQ process number                      & 4       & -- \\
17 & Downlink assignment index                & 0, 2, 4 & Cell group $\rightarrow$ pdsch-HARQ-ACK-Codebook, sCellToAddModList \\
18 & TPC command for scheduled PUCCH          & 2       & -- \\
19 & PUCCH resource indicator                 & 3       & -- \\
20 & \mbox{PDSCH-to-HARQ\_feedback} timing indicator & 0--3 & PUCCH $\rightarrow$ dl-DataToUL-ACK \\
21 & Antenna port(s)                          & 4--6    & PDSCH $\rightarrow$ dmrs-Type, maxLength \\
22 & Transmission configuration indication    & 0, 3    & PDCCH $\rightarrow$ tci-PresentInDCI \\
23 & SRS request                              & 2, 3    & Cell $\rightarrow$ supplementaryUplink \\
24 & CBG transmission information             & 0--8    & PDSCH $\rightarrow$ maxCodeBlockGroupsPerTransportBlock \\
25 & CBG flushing out information             & 0, 1    & PDSCH $\rightarrow$ codeBlockGroupFlushIndicator \\
26 & DMRS sequence initialization             & 1       & -- \\
\bottomrule
\end{tabular}
\begin{tablenotes}[flushleft]
\scriptsize
\item[] Cell: ServingCellConfig;
Cell group: (Physical)CellGroupConfig;
``--'': fixed size;
Var.: variable size;
VRB: virtual resource block;
TB: transport block;
HARQ: hybrid automatic repeat request;
TPC: transmit power control;
CBG: code block group.
\end{tablenotes}
\end{threeparttable}
\end{table}

\section{UE Capability Dataset}
\label{sec:app_ue_caps}
This appendix lists the 21 COTS UE capabilities we replay through the OAI UE~(\autoref{tab:app_cots}) and the capability fields we mutate to generate the 83 variants~(\autoref{tab:app_cap_fields}).

\begin{table}[t]
\centering
\scriptsize
\setlength{\tabcolsep}{4pt}
\begin{threeparttable}
\caption{COTS UEs used to collect UE capabilities.}
\label{tab:app_cots}
\begin{tabular}{@{}rlll@{}}
\toprule
\# & Phone & Baseband & Vendor \\
\midrule
1 & Huawei Pura 90 & Kirin 9010s & HiSilicon \\
2 & Xiaomi POCO M7 Pro & Dimensity 7025-Ultra & MediaTek \\
3 & Xiaomi POCO X8 Pro Max & Dimensity 9500s & MediaTek \\
4 & Google Pixel 8 Pro & Exynos 5300 & Samsung LSI \\
5 & Google Pixel 9 Pro XL & Exynos 5400 & Samsung LSI \\
6 & Xiaomi Redmi K40 Gaming & Dimensity 1200 & MediaTek \\
7 & Xiaomi Redmi Note 13 Pro 5G & Snapdragon 7s Gen 2 & Qualcomm \\
8 & Xiaomi Redmi Note 9T 5G & Dimensity 800U & MediaTek \\
9 & Samsung Galaxy A26 5G & Exynos 1380 & Samsung LSI \\
10 & Samsung Galaxy S20 5G & Snapdragon X55 & Qualcomm \\
11 & Samsung Galaxy S22+ & Snapdragon X65 & Qualcomm \\
12 & Samsung Galaxy S24 Ultra & Snapdragon X75 & Qualcomm \\
13 & Samsung Galaxy Z Flip4 & Snapdragon X65 & Qualcomm \\
14 & Samsung Galaxy Z Flip6 & Snapdragon X75 & Qualcomm \\
15 & Xiaomi 14T & Dimensity 8300-Ultra & MediaTek \\
16 & Xiaomi 15 Ultra & Snapdragon X80 & Qualcomm \\
17 & Xiaomi 17T & Dimensity 8500-Ultra & MediaTek \\
18 & Xiaomi Redmi Turbo 4 & Dimensity 8400-Ultra & MediaTek \\
19 & Apple iPhone 16 & Snapdragon X71 & Qualcomm \\
20 & Apple iPhone 16e & Apple C1 & Apple \\
21 & Apple iPhone 17e & Apple C1X & Apple \\
\midrule
\multicolumn{4}{l}{Total: 5 phone vendors, 5 baseband vendors} \\
\bottomrule
\end{tabular}
\end{threeparttable}
\end{table}

\begin{table*}[p]
\centering
\begin{threeparttable}
\scriptsize
\setlength{\tabcolsep}{3pt}
\caption{Mutated UE capability fields~\cite{3gpp38306}.}
\label{tab:app_cap_fields}
\begin{tabular}{p{1.1cm} p{5.6cm} p{2.4cm} p{2.2cm} c p{4.0cm}}
\toprule
Group & Capability field & Meaning & Mutation & Changes & Resulting RRC Reconfiguration change \\
\midrule
\multicolumn{6}{l}{\textit{Fields that change the RRC Reconfiguration}} \\
\midrule
BWP & \texttt{bwp-SameNumerology} & up to 2/4 BWPs, one numerology & del; upto4 $\to$ upto2; add & K, U & K: needed (with the next two) for BWP2 and the $k_0{=}6$ row. U: absent $\to$ one BWP \\
BWP & \texttt{bwp-SwitchingDelay} & BWP switching delay type & del; add type2 & K & needed for BWP2 \\
Cross-slot & \texttt{dl-SchedulingOffset-PDSCH-TypeA} & cross-slot ($k_0{>}0$) PDSCH type A & del/add one or both copies & K & both copies needed for BWP2 \\
RA & \texttt{dynamicSwitchRA-Type0-1-PDSCH} & dynamic PDSCH RA Type 0/1 & del & K, U & RA dynamicSwitch $\to$ Type 1 \\
DL MIMO & \texttt{maxNumberMIMO-LayersPDSCH} & max PDSCH layers per CC & 4 $\to$ 2 / 8 & K & PDSCH max layers 2/8; RI restriction \\
UL MIMO & \texttt{maxNumberMIMO-LayersCB-PUSCH} & max codebook PUSCH layers & 1 $\leftrightarrow$ 2; $\to$ 4 & K & max rank ($\le$2); 2 layers with 1-port SRS $\to$ UE dropped \\
UL MIMO & \texttt{maxNumberSRS-Ports-PerResource} & max ports per SRS resource & n1 / n2 / n4 & K & 1-port / 2-port SRS ($\le$2) \\
UL MIMO & \texttt{maxNumberSRS-ResourcePerSet} & SRS resources per codebook set & 1 $\to$ 2 & K (bundle) & 2 SRS resources per set \\
UL MIMO & \texttt{supportedSRS-TxPortSwitch} & SRS antenna switching & t2r4 / t1r4 / none & K & antenna-switching SRS set with 2/4/no resources \\
UL MIMO & \texttt{pusch-TransCoherence} & coherence of Tx ports & $\to$ partial / full & K & full $\to$ full codebook subset \\
UL MIMO & \texttt{ul-FullPwrMode-r16} & full-power non-coherent UL & add & K (bundle) & part of the 2-Tx bundle \\
Modulation & \texttt{pdsch-256QAM-FR1} & 256QAM PDSCH & del & K, U & PDSCH 64QAM MCS table; CQI table1 \\
Modulation & \texttt{pusch-256QAM} & 256QAM PUSCH & del & K, U & PUSCH 64QAM MCS table \\
Modulation & \texttt{supportedModulationOrderDL/UL} & max modulation order & qam256 $\to$ qam64; del; add & U & as deleting both 256QAM fields \\
CSI & \texttt{maxNumberTxPortsPerResource} & max CSI-RS ports (Type I) & p32 $\to$ p16 / p8 & U & 8-port NZP-CSI-RS removed \\
CSI & \texttt{simultaneousCSI-ReportsPerCC} & simultaneous CSI reports & $\to$ 1 / 8 & K & 1 $\to$ CSI-IM and CSI NZP-CSI-RS removed \\
Beam & \texttt{maxNumberNonGroupBeamReporting} & beams per report & n4 $\to$ n1 & K & reported RS n2 $\to$ n1 \\
DMRS & \texttt{supportedDMRS-TypeDL/UL} & DMRS types 1 and 2 & $\to$ type1 & K & PDSCH DMRS type2 $\to$ type1 \\
PUCCH & \texttt{mux-SR-HARQ-ACK-CSI-PUCCH-OncePerSlot} & SR+HARQ-ACK+CSI on one PUCCH & del & K & PUCCH F3 simultaneous HARQ-ACK/CSI removed \\
Proc.\ time & \texttt{pdsch/pusch-ProcessingType2} & processing capability 2 & add & U & processing type 2 enabled (UE timing only) \\
CA & \texttt{supportedBandCombinationList} & CA band combinations & add A+A, A/A, D/A; C/C; keep only C or A+A & U & A+A $\to$ SCell (133 PRB), \texttt{reportTriggerSize} +1 \\
Bundle & 2-Tx: CB-PUSCH 2 layers, 2-port SRS, 2 SRS/set, t2r4, full power & second Tx chain & bundle & K & rank 2, 2-port SRS, antenna-switching set; with full coherence $\to$ UE dropped \\
Bundle & floor: 64QAM, 2 DL layers, DMRS type1, no extra DMRS, semi-static HARQ & weakest plausible UE & bundle & K & sum of the single edits \\
Bundle & ceiling: 8 DL layers, 4-layer CB-PUSCH, 4-port SRS, full coherence & strongest plausible UE & bundle & K & 8 DL layers; full codebook subset \\
\midrule
\multicolumn{6}{l}{\textit{Fields with no change in either network}} \\
\midrule
Cross-slot & \multicolumn{5}{p{15.6cm}}{\texttt{dl-SchedulingOffset-PDSCH-TypeB}, \texttt{ul-SchedulingOffset} (K)} \\
RA, UL MIMO & \multicolumn{5}{p{15.6cm}}{\texttt{dynamicSwitchRA-Type0-1-PUSCH}, \texttt{ra-Type0-PUSCH}, non-codebook PUSCH fields (\texttt{maxNumberMIMO-LayersNonCB-PUSCH}, \texttt{maxNumberSRS-ResourcePerSet}, \texttt{maxNumberSimultaneousSRS-ResourceTx})} \\
CSI, beam & \multicolumn{5}{p{15.6cm}}{\texttt{maxNumberAperiodicCSI-triggeringStatePerCC}, \texttt{maxNumberPeriodicCSI-PerBWP-ForCSI-Report}, Type I single-panel \texttt{modes} (U), \texttt{maxNumberActiveTCI-PerBWP}, \texttt{maxNumberCSI-RS-SSB-CBD} (K)} \\
PDCCH, DMRS, PDSCH & \multicolumn{5}{p{15.6cm}}{\texttt{precoderGranularityCORESET} (U), \texttt{twoFL-DMRS} and additional-DMRS fields, \texttt{dynamicPRB-BundlingDL}, \texttt{rateMatchingResrcSetSemi-Static}} \\
UCI, HARQ, PUCCH & \multicolumn{5}{p{15.6cm}}{\texttt{dynamicBetaOffsetInd-HARQ-ACK-CSI}, dynamic/semi-static HARQ-ACK codebook, \texttt{pucch-F2/F3/F4-WithFH}} \\
Coverage & \multicolumn{5}{p{15.6cm}}{\texttt{pusch-HalfPi-BPSK}, \texttt{pucch-F3-4-HalfPi-BPSK}, \texttt{pusch-RepetitionTypeA-r16}, \texttt{tb-ProcessingMultiSlotPUSCH-r17}} \\
Other & \multicolumn{5}{p{15.6cm}}{Rel-16 DCI 0\_2/1\_2 and URLLC fields (U), \texttt{logicalChannelSR-DelayTimer} (K), \texttt{accessStratumRelease}, \texttt{channelBW-90mhz}} \\
\bottomrule
\end{tabular}
\begin{tablenotes}[flushleft]
\scriptsize
\item[] Mutation: del = delete; add = insert with a value copied from a phone that has it; otherwise, change the value.
Changes: networks whose RRC Reconfiguration changed.
\item[] $\ast$
CC: component carrier;
CSI-IM: channel state information interference measurement;
RS: reference signal;
SR: scheduling request;
ACK: acknowledgment;
UCI: uplink control information;
URLLC: ultra-reliable low-latency communications.
F2/F3/F4 denote PUCCH formats 2/3/4.
\end{tablenotes}
\end{threeparttable}
\end{table*}

\section{RRC Reconfiguration Profiles}
\label{sec:app_profiles}

This appendix lists the RRC Reconfiguration profiles that the replayed capabilities produce in Network~K~(\autoref{tab:app_profiles_k}) and Network~U~(\autoref{tab:app_profiles_u}).
A profile is the RRC Reconfiguration the gNB sends after registration, with per-connection values masked (PUCCH PRB offsets, scrambling IDs, SRS positions, and measurement IDs).
DCI sizes are computed from each profile following TS~38.212~\S7.3.1~\cite{3gpp38212}.

\begin{table*}[t]
\centering
\begin{threeparttable}
\scriptsize
\setlength{\tabcolsep}{3pt}
\caption{RRC Reconfiguration profiles of Network~K~\cite{3gpp-5g-rrc}.}
\label{tab:app_profiles_k}
\begin{tabular}{l p{4.1cm} p{4.6cm} c p{5.2cm}}
\toprule
ID & Source & RRC change vs.\ K02 & DCI 1\_1/0\_1 & Effect on decoding \\
\midrule
K01 & V: Z Flip4, DMRS-TypeDL $\to$ type1 & DMRS type2 $\to$ type1 & 51/44 & 1\_1 antenna ports 5 $\to$ 4 bits; DMRS type~1 RE pattern \\
K02 & COTS: 4 phones & -- (reference) & 52/44 & -- \\
K03 & V: Z Flip4, SRS ports n1 $\to$ n4 & 2-port SRS, rank 1, non-coherent & 52/45 & 0\_1 precoding field 0 $\to$ 1 bit \\
K04 & V: Pixel 8 Pro w/o BWP2; Z Flip4 2-Tx bundle & 2-port SRS, rank 2 & 52/46 & precoding 0 $\to$ 2 bits; up to 2 PUSCH layers \\
K05 & V: Pura 90, del.\ dynamicSwitchRA-PDSCH & 2-port SRS, rank 1, full codebook subset; NZP-CSI-RS 4 $\to$ 2 & 52/47 & precoding 0 $\to$ 3 bits; PDSCH rate matching around 2 CSI-RS \\
K06 & V: Pixel 8 Pro, DMRS-TypeDL $\to$ type1 & BWP2; DMRS type1; 2-port SRS, rank 2 & 52/47 (51/46) & +1 BWP-indicator bit; cross-slot TDRA row; BWP switch changes all sizes \\
K07 & V: Z Flip4 ceiling & 2-port SRS, rank 2, full subset; 8 PDSCH layers & 52/48 & precoding 0 $\to$ 4 bits \\
K08 & COTS: iPhone 16e, 17e, S24 Ultra & BWP2 & 53/45 (52/44) & +1 BWP bit; FDRA 16 $\to$ 15 bits in BWP2; cross-slot PDSCH; second CORESET \\
K09 & V: Pixel 8 Pro, CB-PUSCH 2 $\to$ 1 layer & BWP2; 2-port SRS, rank 1 & 53/46 (52/45) & K08 + 1-bit precoding \\
K10 & COTS: Pixel 8 Pro, Pixel 9 Pro XL & BWP2; 2-port SRS, rank 2 & 53/47 (52/46) & K08 + 2-bit precoding; up to 2 PUSCH layers \\
K11 & V: Pixel 8 Pro ceiling & BWP2; 2-port SRS, rank 2, full subset; 8 PDSCH layers & 53/49 (52/48) & K08 + 4-bit precoding \\
K12 & COTS: Pura 90 & resource allocation dynamicSwitch; 2-port SRS, rank 1, full subset; NZP-CSI-RS 2 & 55/47 & 1\_1 FDRA 16 $\to$ 19 bits, MSB selects Type 0 or 1; 3-bit precoding \\
K13 & V: Pura 90, CB-PUSCH $\to$ 4 layers & K12 with rank 2 & 55/48 & K12 + 4-bit precoding \\
K14 & COTS: POCO M7 Pro, X8 Pro Max, K40 Gaming, Note 9T, 14T & BWP2; resource allocation dynamicSwitch & 56/45 (53/44) & K08 + dynamic Type 0/1 FDRA \\
K15 & V: Pura 90, add dl-SchedulingOffset & BWP2; dynamicSwitch; 2-port SRS, rank 1, full subset & 56/48 (53/47) & K14 + 3-bit precoding \\
K16 & V: Z Flip4 floor & K01 + 64QAM MCS tables, CQI table1, 2 PDSCH layers & 51/44 & same size as K01; wrong MCS table fails every TB; LBRM \\
K17 & V: Z Flip4, del.\ both 256QAM & 64QAM MCS tables; CQI table1 & 52/44 & MCS table only \\
K18 & V: Pixel 8 Pro floor & K06 + 64QAM tables; 2 PDSCH layers & 52/47 (51/46) & K06 sizes; MCS table; LBRM \\
K19 & V: Pixel 8 Pro, del.\ both 256QAM & K10 + 64QAM tables & 53/47 (52/46) & K10 sizes; MCS table \\
K20 & V: Pura 90, del.\ both 256QAM & K12 + 64QAM tables & 55/47 & K12 sizes; MCS table \\
K21 & V: Z Flip4, PDSCH 4 $\to$ 2 layers & 2 PDSCH layers; RI restriction & 52/44 & LBRM buffer size; smaller RI field in CSI \\
\bottomrule
\end{tabular}
\begin{tablenotes}[flushleft]
\scriptsize
\item[] Reference profile K02 (COTS: Galaxy Z Flip4, Z Flip6, iPhone 16, Redmi Note 13 Pro): one 273-PRB BWP; 12-row PDSCH TDRA, all $k_0{=}0$; RA Type~1; DMRS type~2; 256QAM MCS tables; 4 PDSCH layers; codebook PUSCH with a 1-port SRS and rank~1; 4 NZP-CSI-RS; DCI 1\_1/0\_1 = 52/44 bits.
``BWP2'' adds a second 217-PRB BWP with its own CORESET and a 13th PDSCH TDRA row with $k_0{=}6$.
Source: COTS = unmodified capability of the listed phones; V = reached only by a variant (seed phone and mutation given).
DCI sizes in parentheses are for BWP2.
\item[] $\ast$
TDRA: time-domain resource assignment;
QAM: quadrature amplitude modulation;
RA: resource allocation;
RE: resource element;
Tx: transmit;
CB-PUSCH: codebook-based PUSCH;
MSB: most significant bit;
CQI: channel quality indicator;
LBRM: limited buffer rate matching;
RI: rank indicator.
\end{tablenotes}
\end{threeparttable}
\end{table*}

\begin{table*}[t]
\centering
\begin{threeparttable}
\scriptsize
\setlength{\tabcolsep}{3pt}
\caption{RRC Reconfiguration profiles of Network~U~\cite{3gpp-5g-rrc}.}
\label{tab:app_profiles_u}
\begin{tabular}{l p{4.6cm} p{4.6cm} c p{5.0cm}}
\toprule
ID & Source & RRC change vs.\ U01 & DCI 1\_1/0\_1 & Effect on decoding \\
\midrule
U01 & COTS: Pixel 8 Pro & -- (reference) & 52/45 & -- \\
U02 & COTS: Pixel 9 Pro XL & \texttt{t312} timer added & 52/45 & none \\
U03 & COTS: POCO M7 Pro, K40 Gaming, Note 9T, 14T, 17T & resource allocation dynamicSwitch; SDAP header DL & 55/45 & 1\_1 FDRA 16 $\to$ 19 bits; 1-byte SDAP header before every DL SDU \\
U04 & COTS: S24 Ultra, iPhone 16e, 17e & overheating assistance removed & 52/45 & none \\
U05 & COTS: Pura 90 & dynamicSwitch; 8-port CSI-RS removed & 55/45 & U03 layout; no rate matching around the 8-port CSI-RS; no SDAP header \\
U06 & COTS: Note 13 Pro, 15 Ultra, Z Flip4 (also S20, S22+, Z Flip6, iPhone 16) & one BWP; 7 common TDRA rows, all $k_0{=}0$; \texttt{reportTriggerSize} 1 & 50/43 & no BWP indicator; TDRA 4 $\to$ 3 bits; CSI request 2 $\to$ 1 bit \\
U07 & V: Pixel 8 Pro, add A+A & SCell (133 PRB); \texttt{reportTriggerSize} 3 & 54/46 & DAI 2 $\to$ 4 bits; CSI request 2 $\to$ 3 bits; SCell has its own DCI sizes \\
U08 & V: POCO M7 Pro, add A+A & U07 + dynamicSwitch + SDAP header & 57/46 & U07 + dynamic FDRA \\
U09 & V: Pixel 8 Pro, del.\ both 256QAM or modulation order qam64 & 64QAM MCS tables DL and UL; CQI table1 & 52/45 & TBS and modulation of every grant \\
U10 & V: Pixel 8 Pro, del.\ pusch-256QAM & 64QAM PUSCH tables & 52/45 & PUSCH MCS table only \\
U11 & V: Pixel 8 Pro, CSI-RS ports p32 $\to$ p8 & 8-port CSI-RS removed & 52/45 & no rate matching around the 8-port CSI-RS \\
U12 & V: Pixel 8 Pro, CSI-RS ports p32 $\to$ p16 & U11 + periodic SSB report & 52/45 & as U11 \\
U13 & V: Xiaomi 14T, reduced CSI & U03 + periodic SSB report & 55/45 & as U03 \\
U14 & V: POCO M7 Pro, del.\ both 256QAM & U03 + U09 & 55/45 & U03 layout; 64QAM DL and UL \\
U15 & V: POCO M7 Pro, del.\ pusch-256QAM & U03 + U10 & 55/45 & U03 layout; 64QAM PUSCH \\
U16 & V: POCO M7 Pro, del.\ bwp-SameNumerology & one BWP + dynamicSwitch & 53/43 & U06 layout with 19-bit FDRA \\
U17 & V: Note 13 Pro, add A+A & one BWP; SCell; \texttt{reportTriggerSize} 2 & 52/44 & U06 + 4-bit DAI + 2-bit CSI request \\
U18 & V: POCO M7 Pro, del.\ bwp-SameNumerology, add A+A & one BWP; dynamicSwitch; SCell; \texttt{reportTriggerSize} 2 & 55/44 & U17 with 19-bit FDRA \\
\bottomrule
\end{tabular}
\begin{tablenotes}[flushleft]
\scriptsize
\item[] Reference profile U01 (COTS: Pixel 8 Pro): two 273-PRB BWPs; 10-row PDSCH TDRA, rows 8--10 with $k_0{=}5/6/7$; RA Type~1; 256QAM MCS tables; DFT-s-OFDM PUSCH with a 1-port SRS and rank~1; CQI table2; 8-port and 32-port NZP-CSI-RS; CSI \texttt{reportTriggerSize}~2; DCI 1\_1/0\_1 = 52/45 bits.
A+A: intra-band non-contiguous CA band combination.
\item[] $\ast$
RA: resource allocation;
DFT-s-OFDM: discrete Fourier transform-spread orthogonal
frequency-division multiplexing;
SCell: secondary cell;
DAI: downlink assignment index;
SDU: service data unit;
TBS: transport block size.
\end{tablenotes}
\end{threeparttable}
\end{table*}

\section{Derivation of Syndrome-Guided C-RNTI Recovery}
\label{sec:app_syndrome}
This appendix details the C-RNTI recovery pipeline of \autoref{subsubsec:syndrome_rnti_recovery} step by step, extending it from the length-$N$ polar codeword to the rate-matched PDCCH bits that \sys receives.
All binary operations are over $\mathrm{GF}(2)$.

\nibf{Step~1: PDCCH encoding and RNTI-dependent scrambling.}
The C-RNTI enters the PDCCH at two points.
First, the gNB appends a 24-bit CRC to the $A$-bit DCI payload $\mathbf{a}$ and masks the last 16 CRC bits with the C-RNTI $r^\star$~\cite{3gpp38212}:
\begin{equation}
    \mathbf{p}=C(\mathbf{a})\oplus(\mathbf{0}_{8}\,\Vert\,r^\star),
    \label{eq:app_crc_mask}
\end{equation}
where $C(\cdot)$ computes the CRC and $\Vert$ denotes concatenation.
The $K=A+24$ bits $(\mathbf{a},\mathbf{p})$ are placed on the information positions of $\mathbf{u}$ and polar-encoded into $\mathbf{x}=\mathbf{u}G_N$, with $N\leq512$ for the PDCCH.
Rate matching then adapts the $N$ codeword bits to the $E=108L$ bits carried by $L$ CCEs, producing $\mathbf{x}_E$.
Second, the gNB scrambles all $E$ coded bits with a sequence $\mathbf{c}_E(r^\star)$ initialized by Eq.~\eqref{eq:pdcch_scrambling_init}.
The CRC mask thus acts on information bits before polar encoding, whereas scrambling acts on coded bits after rate matching.
\sys receives hard decisions
\begin{equation}
    \hat{\mathbf{b}}_E=\mathbf{x}_E\oplus\mathbf{c}_E(r^\star)\oplus\mathbf{e}_E,
    \label{eq:app_received_E}
\end{equation}
where $\mathbf{e}_E$ represents hard-decision errors.

\nibf{Step~2: Frozen-bit syndrome in the codeword domain.}
The CRC mask changes only information bits, and the frozen positions $\mathcal{F}$ of $\mathbf{u}$ remain zero for every C-RNTI.
Every PDCCH codeword therefore satisfies $f(\mathbf{x})=\mathbf{0}$ under Eq.~\eqref{eq:frozen_syndrome}, regardless of the payload and the C-RNTI.
The C-RNTI affects the syndrome only through scrambling.
If the received bits mapped one-to-one to the length-$N$ codeword ($E=N$), this would yield $H=S(r^\star)\oplus f(\mathbf{e})$ in Eq.~\eqref{eq:syndrome_separation} and the hypothesis test in Eq.~\eqref{eq:syndrome_lookup}.
Since $E\neq N$ in general, \sys first undoes rate matching to map the received hard decisions back to codeword positions (Step~3) and then evaluates the syndrome on those positions (Step~4).

\nibf{Step~3: Inverse rate matching.}
After sub-block interleaving $\pi$, transmitted bit $k$ carries codeword position $p_k$ (TS~38.212~\S5.4.1)~\cite{3gpp38212}:
\begin{equation}
    p_k=
    \begin{cases}
        \pi(k\bmod N), & \text{repetition},\\
        \pi(k+N-E), & \text{puncturing},\\
        \pi(k), & \text{shortening}.
    \end{cases}
    \label{eq:app_rate_matching}
\end{equation}
Repetition applies when $E\geq N$; otherwise, the gNB punctures when $K/E\leq 7/16$ and shortens when $K/E>7/16$~\cite{3gpp38212}.
Since $E=108L$ exceeds $N\leq512$ for $L\geq8$, aggregation levels 8 and 16 always use repetition, whereas lower aggregation levels puncture or shorten depending on $K$.
The three modes affect the receiver differently.
Repetition provides several observations of the same codeword bit.
Shortening drops positions that the encoder freezes to zero, so the receiver knows their values.
Puncturing drops positions whose values the receiver does not know.
\sys inverts Eq.~\eqref{eq:app_rate_matching} with a mapping $D:\{0,1\}^E\rightarrow\{0,1\}^N$ that keeps the first copy of each repeated position and fills dropped positions with zeros.
$D$ only selects and places bits, so it is linear.
Let $\mathcal{P}$ denote the punctured positions, which is empty under repetition and shortening.
For a valid codeword, $D(\mathbf{x}_E)$ equals $\mathbf{x}$ at every position outside $\mathcal{P}$.

\nibf{Step~4: Rate-matching-aware syndrome and signatures.}
Each frozen-bit check $j\in\mathcal{F}$ is the XOR of the codeword positions
\begin{equation}
    \mathcal{S}_j=\bigl\{\,i \mid (G_N^{-1})_{i,j}=1\,\bigr\}.
    \label{eq:app_check_support}
\end{equation}
A check is clean if it reads no punctured position, and \sys keeps only clean checks:
\begin{equation}
    \mathcal{J}=\bigl\{\,j\in\mathcal{F} \mid \mathcal{S}_j\cap\mathcal{P}=\varnothing\,\bigr\}.
    \label{eq:app_clean_checks}
\end{equation}
Under repetition and shortening, $\mathcal{P}$ is empty and every frozen check is clean.
Among the clean checks, \sys prefers checks with small $|\mathcal{S}_j|$, because a check that XORs more bits is more likely to be flipped by a hard-decision error.
\sys then defines the rate-matching-aware syndrome
\begin{equation}
    f_E(\mathbf{z})=\bigl(D(\mathbf{z})\,G_N^{-1}\bigr)_{\mathcal{J}},
    \label{eq:app_syndrome_E}
\end{equation}
and applies the same $f_E$ to the received bits and to every scrambling sequence, giving $H=f_E(\hat{\mathbf{b}}_E)$ and $S(r)=f_E(\mathbf{c}_E(r))$.
Because each check in $\mathcal{J}$ reads only positions where $D(\mathbf{x}_E)$ equals $\mathbf{x}$, it returns the same value as on $\mathbf{x}$, which is zero.
Hence $f_E(\mathbf{x}_E)=\mathbf{0}$, and the linearity of $f_E$ applied to Eq.~\eqref{eq:app_received_E} gives
\begin{equation}
    H=S(r^\star)\oplus f_E(\mathbf{e}_E),
    \label{eq:app_rm_syndrome}
\end{equation}
which has the same form as Eq.~\eqref{eq:syndrome_separation}.

\nibf{Step~5: Signature precomputation and candidate ranking.}
Signatures depend only on the DCI size, the aggregation level, the scrambling identity, and the RNTI hypothesis, not on the received signal.
\sys therefore precomputes one signature table per combination of DCI size, aggregation level, and scrambling identity, and computes $H$ once per received candidate.
Because the modulo in Eq.~\eqref{eq:pdcch_scrambling_init} removes the most significant bit of $r$, each table holds $2^{15}$ signatures, each shared by the equivalence class $\{r,\,r\oplus 2^{15}\}$.
\sys keeps a class only if one of its two RNTIs has a search space that can place a candidate at the observed CCE location~(Eq.~\eqref{eq:rnti_cce_mapping}, \autoref{subsubsec:rnti_candidate_reduction}).
For each remaining class, it computes the Hamming distance $d(r)=\operatorname{wt}(H\oplus S(r))$ with an XOR and a population count.
For the correct class, $d(r^\star)=\operatorname{wt}(f_E(\mathbf{e}_E))$ counts only the checks flipped by hard-decision errors.
For an incorrect class, $H\oplus S(r)$ also contains $S(r^\star)\oplus S(r)$, which typically flips about half of the checks.
Because noise can still push the correct class below others, \sys shortlists the 16 classes with the smallest distances instead of only the closest one.

\nibf{Step~6: Verification by full polar decoding.}
For each shortlisted class, \sys descrambles the received LLRs with $\mathbf{c}_E(r)$, undoes rate matching, and runs a full soft-decision polar decoder, such as successive cancellation~(SC) decoding.
Unlike the syndrome, the decoder uses the reliability of every received bit and can therefore correct the errors that perturbed $H$.
\sys accepts a class only if the decoded CRC is valid after removing the C-RNTI mask.
Because the mask covers all 16 bits of the C-RNTI, the decoded CRC also reveals the most significant bit that the scrambling hides, yielding the full C-RNTI.
\sys further validates each accepted DCI with the configuration profile~(\autoref{subsubsec:false_positive_dci}).

\end{document}